\documentclass[twocolumn,superscriptaddress,floatfix,preprintnumbers, nofootinbib,hyperref]{revtex4} 
\usepackage[colorlinks=true,breaklinks=true]{hyperref}
\usepackage{cancel}
\usepackage[normalem]{ulem}
\usepackage[utf8]{inputenc}
\hypersetup{allcolors=[rgb]{0.0 0.0 0.6},linkcolor=[rgb]{0.75 0.05 0.05}}
\usepackage{amsmath,amssymb}
\usepackage{epsfig}  
\usepackage{graphicx}   
\usepackage{slashed}             
\usepackage{url}
\usepackage{color}
\usepackage{multirow}
\usepackage[dvipsnames]{xcolor}
\usepackage{letltxmacro}
\usepackage{amssymb}

\LetLtxMacro{\oldcite}{\cite}
\renewcommand{\cite}[1]{\mbox{\oldcite{#1}}}

\newcommand{\gag}{g_{a\gamma}}

\DeclareMathOperator{\sinc}{sinc}

\allowdisplaybreaks

\bibpunct{[}{]}{,}{n}{}{,}

\begin{document}

\title{Turbulent axion-photon conversions in the Milky-Way}
\author{Pierluca Carenza}
\email{pierluca.carenza@ba.infn.it}
\affiliation{Dipartimento Interateneo di Fisica ``Michelangelo Merlin'', Via Amendola 173, 70126 Bari, Italy.}
\affiliation{Istituto Nazionale di Fisica Nucleare - Sezione di Bari,
Via Orabona 4, 70126 Bari, Italy.}             

\author{Carmelo Evoli}
\email{carmelo.evoli@gssi.it}
\affiliation{Gran Sasso Science Institute (GSSI), Viale Francesco Crispi 7, 67100 L'Aquila, Italy}
\affiliation{INFN-Laboratori Nazionali del Gran Sasso (LNGS),  via G. Acitelli 22, 67100 Assergi (AQ), Italy}

\author{Maurizio Giannotti}
\email{MGiannotti@barry.edu}
\affiliation{Physical Sciences, Barry University, 11300 NE 2nd Ave., Miami Shores, FL 33161, USA.}

\author{Alessandro Mirizzi}
\email{alessandro.mirizzi@ba.infn.it }
\affiliation{Dipartimento Interateneo di Fisica ``Michelangelo Merlin'', Via Amendola 173, 70126 Bari, Italy.}
\affiliation{Istituto Nazionale di Fisica Nucleare - Sezione di Bari,
Via Orabona 4, 70126 Bari, Italy.}

\author{Daniele Montanino}
\email{daniele.montanino@le.infn.it }
\affiliation{Dipartimento di Matematica e Fisica ``Ennio De Giorgi'', Universit\`a del Salento, Via Arnesano, 73100 Lecce, Italy}
\affiliation{Istituto Nazionale di Fisica Nucleare - Sezione di Lecce,
Via Arnesano, 73100 Lecce, Italy.}       


\begin{abstract}
The Milky-Way magnetic field can trigger conversions between photons and axion-like particles (ALPs), leading to peculiar features on the observable photon spectra. 
Previous studies considered only the regular component  of the magnetic field. 
However, observations consistently show the existence of an additional turbulent component, with a similar strength and correlated on a scale of a few 10$\,$pc.
We investigate the impact of the turbulent magnetic field on the ALP-photon conversions, characterizing the effects numerically and analytically. 
We show that the turbulent magnetic field can change the conversion probability by up to a factor of two and may lead to observable irregularities in the observable photon spectra from different astrophysical sources.
 \end{abstract}

\maketitle

\section{Introduction}

Conversions between photons and axion-like particles (ALPs)  in cosmic magnetic fields are the subject of intense investigations (see, e.g.,~\cite{DeAngelis:2007wiw,Mirizzi:2006zy,Mirizzi:2007hr, Hooper:2007bq,Simet:2007sa,Day:2015xea,Galanti:2018upl,Berg:2016ese,Majumdar:2018sbv,marsh}).
If ALPs exist, the analysis of the gamma-ray spectra from Galactic and extragalactic sources may reveal valuable information on their coupling with photons $g_{a\gamma}$ and mass $m_a$.
Several specific cases have been studied so far.
Conversions of photons into ALPs in cosmic magnetic fields of Galactic or extra-galactic origin may imprint peculiar deformations on the spectra of very-high-energy photons from faraway sources, such as blazars, active galactic nuclei, pulsars and galaxy clusters (see, e.g.~\cite{Meyer:2013pny,Berg:2016ese, Galanti:2018upl, Majumdar:2018sbv, Buehler:2020qsn,Montanino:2017ara}), and alter the polarization of X-rays in a measurable way~\cite{Bassan:2010ya,Mena:2011xj}.
Dimming effects on the photons from Supernovae Ia induced by photons oscillating into ALPs and viceversa in the intergalactic magnetic field have also been characterized~\cite{Csaki:2001yk,Mirizzi:2005ng,Avgoustidis:2010ju}. 
Finally, ALPs produced in the hot core of massive stars such as red  supergiants~\cite{Carlson:1995wa,Xiao:2020pra} and core-collapse supernovae~\cite{Brockway:1996yr,Grifols:1996id,Payez:2014xsa}, and converted in the Milky-Way magnetic field, would produce a copious photon flux.

Because of the above considerations, the Milky-Way magnetic field is recognized as a valuable tool to understand the properties of ALPs, triggering ALP-photon conversions for both Galactic and extra-galactic sources (see, e.g.~\cite{Simet:2007sa,Mirizzi:2007hr}).
The Milky-Way magnetic field models adopted in these studies (see, e.g.~\cite{Galanti:2018upl, Majumdar:2018sbv}) make use of state-of-the-art simulations, based on the combined analyses of Faraday rotation measurements of extra-galactic sources and of polarized galactic diffuse radio emission~\cite{Jansson:2012pc,Pshirkov:2011um}.
Typically these simulations resolve the large-scale fields (also called \emph{regular} or coherent) indicating a magnetic field with strength $B\sim {\mathcal O}(1)$ $\mu G$ that is smooth on Galactic scales, usually assumed to follow the spiral arms~\cite{Haverkorn2015}.
However, radio observations of synchrotron emission and polarization reveal also a small-scale component of the magnetic field connected to the turbulent Inter Stellar Medium~(see, e.g., \cite{Reich2004}).
This component is known as the \emph{random} or \emph{turbulent} magnetic field, and has an amplitude comparable to that of the regular magnetic field, but a much shorter correlation length, $l_{\rm corr} \sim {\mathcal O}(10-100)$~pc~\cite{Ohno1993,Haverkorn:2008tb,Malkov:2010yq,Iacobelli:2013fqa}.

The impact of the turbulent component has been neglected in previous investigations, on the basis that the ALP oscillation length would necessarily be much larger than the correlation length of the turbulent field (see, however,~\cite{Mirizzi:2007hr}) and so the effects of the random field would be averaged to zero over the oscillation length.
However, though it is true that in the most interesting cases $l_{\rm osc}$ is much larger than the correlation length $l_{\rm corr}$ of the random field, this condition does not justify neglecting the turbulent component of the magnetic field. 
On the contrary, we will show that in certain cases and depending on the ALP parameters, the turbulent component may contribute as much as the regular component to the oscillation probability.
In general, as we shall see,  the presence of a sizable turbulent component in the Galactic $B$-field leads to peculiar irregularities in the observable 
photon spectra from different astrophysical sources.

The plan of our work is as follows. 
In Section \ref{sec:ALP-to-photon conversions in the Milky-Way}, we present the equations of motion of the ALP-photon mixing in the Galactic magnetic field accounting for the regular and the turbulent component.
We discuss our numerical and analytical approach to characterize the conversions in turbulent $B$-fields and we present different representative examples.
In Section \ref{sec:Effect_photon_spectra}, we show the spectral features induced by the conversion in the turbulent $B$-field observable in photon spectra from different sources.
Finally, in Section \ref{sec:Conclusions} we discuss our results and draw our conclusions. 
There follow two Appendixes, where we give details on our analytical recipe to calculate the ALP-photon conversion probabilities in a regular plus turbulent magnetic field configuration in a perturbative (Appendix A) and non perturbative (Appendix B) regime.

\section{ALP-to-photon conversions in the Milky-Way}
\label{sec:ALP-to-photon conversions in the Milky-Way}

\subsection{Equations of motion}

The Lagrangian describing the ALP-photon system is 
\begin{equation}
{\cal L} = {\cal L}_\gamma +{\cal L}_a + {\cal L}_{a\gamma} \,\ .
\label{eq:lagr}
\end{equation}
The first term in Eq.~\eqref{eq:lagr} is the QED Lagrangian for  photons
\begin{equation}
{\cal L_{\gamma}} = -\frac{1}{4} F_{\mu \nu} \, F^{\mu \nu} 
+ \frac{{\alpha}^2}{90 \, m^4_e} \, \left[ \left(F_{\mu \nu} \, F^{\mu \nu} 
\right)^2 + \frac{7}{4} \left(F_{\mu \nu} \, \tilde F^{\mu \nu} \right)^2 \right]~,
\label{eq:lagr211209}
\end{equation}
where $F_{\mu \nu} \equiv ({\bf E}, {\bf B})$ is the electromagnetic field tensor, $\tilde{F}_{\mu\nu} = \frac{1}{2}\epsilon_{\mu\nu\rho\sigma}F^{\rho\sigma}$ is its dual, $\alpha$ is the fine-structure constant and $m_e$ is the electron mass. 
The second term on the r.h.s. of Eq.~(\ref{eq:lagr211209}) is the Euler-Heisenberg-Weisskopf (HEW) effective Lagrangian~\cite{Raffelt:1987im}, which accounts for the one-loop corrections to classical electrodynamics. 
The Lagrangian for the non-interacting ALP field $a$ in Eq.~(\ref{eq:lagr}) is 
\begin{equation}
{\cal L}_a = \frac{1}{2} \, \partial^{\mu} a \, \partial_{\mu} a - \frac{1}{2} \, m_a^2 \, a^2 \,\ ,
\end{equation}
where $m_a$ is the ALP mass. Finally, the ALP-to-photon interaction in Eq.~(\ref{eq:lagr}) is represented by the following Lagrangian term~\cite{Raffelt:1987im}
\begin{equation}
{\cal L}_{a\gamma}=-\frac{1}{4} \,\gag
F_{\mu\nu}\tilde{F}^{\mu\nu}a=\gag \, {\bf E}\cdot{\bf B}\,a~,
\end{equation}
where $\gag$ is the ALP-photon coupling constant.

We shall consider a monochromatic photon/ALP beam of energy $E$ propagating along the $z$-direction in the presence of an external magnetic field ${\bf B}$. The linearized equations of motion for the photon/ALP system are given by~\cite{Raffelt:1987im}
\begin{equation}
\label{we} 
\left(-i \, \frac{d}{d z} + E +  {\cal M} \right)  \left(\begin{array}{c}A_x (z) \\ A_y (z) \\ a (z) \end{array}\right) = 0~,
\end{equation}
where $A_x (z)$ and $A_y (z)$ are the two photon linear polarization amplitudes along the $x$ and $y$ axis, respectively, $a (z)$ denotes the ALP amplitude and ${\cal M}$ represents the ALP-to-photon mixing matrix. 

The mixing matrix ${\cal M}$ takes a simpler form if we consider the case of a photon beam propagating in a single magnetic domain with an homogeneous field inside. We denote by ${\bf B}_T$ the transverse magnetic field, i.e. its component in the plane normal to the beam direction. We can choose the $y$-axis along ${\bf B}_T$ so that $B_x$ vanishes. Under these simplifying assumptions, the mixing matrix can be written as~\cite{Mirizzi:2006zy} 
\begin{equation}
{\cal M}_0 =   \left(\begin{array}{ccc}
\Delta_{ \perp}  & 0 & 0 \\
0 &  \Delta_{ \parallel}  & \Delta_{a \gamma}  \\
0 & \Delta_{a \gamma} & \Delta_a 
\end{array}\right)~,
\label{eq:massgen}
\end{equation}
whose elements are~\cite{Raffelt:1987im}
\begin{equation}
\Delta_\parallel \equiv \Delta_{\rm pl} + 3.5 \, \Delta_{\rm QED}~,
\end{equation}
\begin{equation}
\Delta_\perp \equiv \Delta_{\rm pl} + 2 \, \Delta_{\rm QED}~,
\end{equation}
\begin{eqnarray}
\Delta_{a\gamma} &\equiv & \frac{1}{2} g_{a\gamma} B_T  \nonumber \\
&\simeq& 1.52 \times 10^{-2} \left(\frac{g_{a\gamma}}{10^{-11} \, {\rm GeV}^{-1}} \right) \nonumber \\
&\times & \left(\frac{B_T}{10^{-6} \,\rm G}\right) {\rm kpc}^{-1}~,
\end{eqnarray}
\begin{eqnarray}
\Delta_a &\equiv& - \frac{m^2}{2E} \nonumber \\
 &\simeq & -7.8 \times 10^{1} \left(\frac{m_a}{10^{-10} \, {\rm eV}}\right)^2 \nonumber \\
 &\times& \left(\frac{E}{{10 \, \rm keV}} \right)^{-1} 
{\rm kpc}^{-1}~,
\end{eqnarray}
with
\begin{eqnarray}
\Delta_{\rm pl} & \equiv & -\frac{\omega^2_{\rm pl}}{2E}  \nonumber \\
& \simeq & - 1.1 \times10^{-1} \left(\frac{E}{{10\, \rm keV}}\right)^{-1}  \nonumber \\
& \times &  \left(\frac{n_e}{10^{-2} \, {\rm cm}^{-3}}\right) {\rm kpc}^{-1}~,
\end{eqnarray}
\begin{eqnarray}
\Delta_{\rm QED} &\equiv& \frac{\alpha E}{45 \pi} \left(\frac{B_T}{B_{\rm cr}} \right)^2  \nonumber \\
& \simeq &  4.1 \times 10^{-14}\left(\frac{E}{{10\, \rm keV}}\right) \nonumber \\
& \times & \left(\frac{B_T}{10^{-6}\,\rm G}\right)^2 {\rm kpc}^{-1}~,
\end{eqnarray}
where $n_e$ is the electron density in the medium, $\omega^2_{\rm pl} = 4 \pi \alpha n_e/m_e$ is the associated plasma frequency, $B_{\rm cr} \equiv m^2_e /e \simeq 4.41 \times 10^{13} \, {\rm G}$ is the critical magnetic field and $e$ denotes the electron charge. 
Concerning the ALP-to-photon coupling $g_{a\gamma}$ we took a representative value below the CAST bound for ultralight ALPs from solar searches, $g_{a\gamma} < 6.6 \times 10^{-11} \, {\rm GeV}^{-1}$~\cite{Anastassopoulos:2017ftl}, while the strength of the magnetic field $B_T$ and the electron density $n_e$ are given by typical conditions in the Milky-Way~\cite{density}.
For the range of energies we are considering in this work, $\Delta_{\rm QED}$ is negligible and will be neglected hereafter.

\subsection{Conversions in the regular magnetic field} 

The ALP-photon oscillation probability takes a particularly simple form in the case of a uniform magnetic field. 
This is a good approximation for the regular magnetic field component, since our ultimate goal is to show the effects of adding a turbulent component over it. Our qualitative results are not modified when this assumption is lifted.

With this assumption, the ALP-photon conversion probability can be calculated analytically.
Considering a photon initially polarized along the $y$ axis, the probability of conversion into an ALP after a distance $r$ reads~\cite{Raffelt:1987im}
\begin{equation}
\label{a16}
P^{(0)}_{a \gamma} ={\rm sin}^2 2 \theta \  {\rm sin}^2
\left( \frac{\Delta_{\rm osc} \, r}{2} \right)~,
\end{equation}
where the mixing angle $\theta$ is given by
\begin{equation}  \label{theta}
\theta = \frac{1}{2} \arctan \left(\frac{ 2 \Delta_{a \gamma}}{\Delta_{\parallel}-\Delta_a} \right)
\end{equation}
and the oscillation wave number is
\begin{equation}
\label{a17}
{\Delta}_{\rm osc} = \left[\left(\Delta_a - \Delta_{\parallel} \right)^2 + 4 \Delta_{a \gamma}^2 \right]^{1/2}~.
\end{equation}

%
\begin{figure}[t!]
\vspace{0.cm}
\includegraphics[width=0.95\columnwidth]{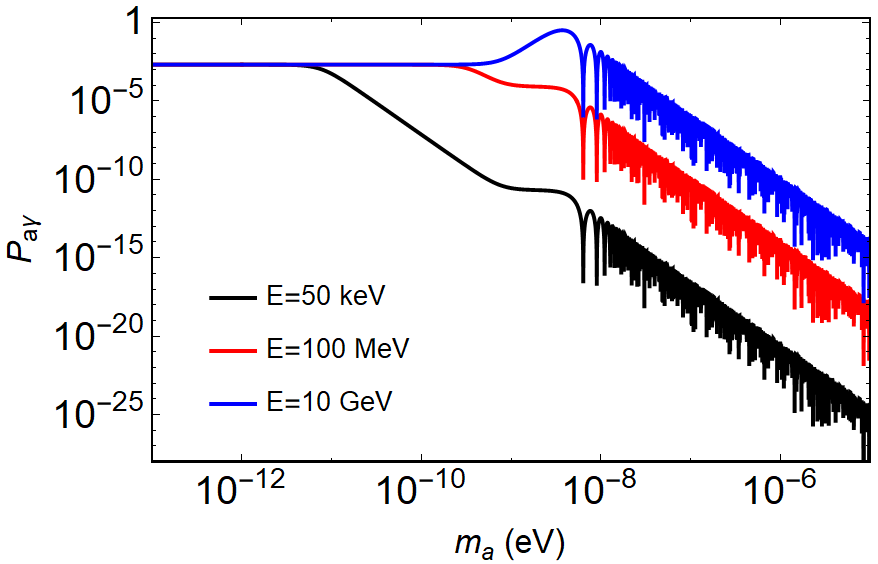}
\caption{Conversion probability $P_{a\gamma}$ for a source at distance $r=1$~kpc
for a value of the regular magnetic field $B_T= 3$ $\mu$G in function of the ALP mass $m_a$. We take 
$g_{a\gamma}=10^{-11}\rm GeV^{-1}$ and 
we consider three representative ALP energies, 
namely
 $E=50$ keV (black curve),
$E=100$ MeV (red curve) and $E=10$ GeV (blue curve).
{The oscillation probability drops approximately as $m_a^{-4}$ when coherence is lost, discussed in the text.} 
}
\label{fig:prob}
\end{figure}

In order to gain more physical intuition, it is convenient to rewrite Eq.~\eqref{a16} as
\begin{align}
P_{a\gamma}^{(0)}=\left( \Delta_{a\gamma}r \right)^2 \sinc^2\left( \Delta_{\rm osc}r /2 \right) \,\ ,
\label{eq:prob0}
\end{align}
where the sinc function is defined as $\sinc x \equiv x^{-1}\sin x$.
The probability in Eq.~(\ref{eq:prob0}) has different limiting regimes, depending on the relative importance of $\Delta_{a\gamma}$ and $|\Delta_a - \Delta_{\parallel}|$ in Eq.~(\ref{a16}).
Specifically, when $\Delta_{a\gamma} \gg |\Delta_a - \Delta_{\parallel}|$,  the probability scales as $\sin^2(\Delta_{a\gamma}r/2)$. However, since 
$\Delta_{a \gamma}r\simeq 1.5\times 10^{-2}g_{11}(B/{\rm \mu G})(r/{\rm kpc})\ll1$ for representative ALP-photon couplings, $g_{11}\equiv g_{a\gamma}/(10^{11}$~GeV$^{-1})\sim 1$, galactic distances, and typical values of the galactic magnetic field, for all practical purposes the oscillation probability in this case reduces to
\begin{equation}
\label{eq:penindep}
P_{a\gamma}^{(0)}= 2.2\times 10^{-4} g_{11}^{2}\left( \dfrac{B_\mathrm{T}}{1~\mu \mathrm{G}}\right)^2 
\left( \dfrac{r}{1\,{\rm kpc}}\right)^2
 \,\ .
\end{equation}
On the other hand, if $\Delta_{a\gamma} \ll |\Delta_a - \Delta_{\parallel}|$
we can conveniently rewrite the conversion probability as  
\begin{equation}
P_{a\gamma}^{(0)} = 2.2\times 10^{-4} g_{11}^{2}\left( \dfrac{B_\mathrm{T}}{1~\mu \mathrm{G}}\right)^2 \left( \dfrac{r}{1\,{\rm kpc}}\right)^2\dfrac{\sin^2 \Phi}{\Phi^2}
\,\ ,
\label{eqa:prob}
\end{equation}
where 
\begin{eqnarray}
\Phi &\simeq \left[ 39\,\left(\dfrac{m_{a}}{10^{-10}\,{\rm eV}}\right)^2-0.060
\left( \dfrac{n_e}{10^{-2}\,{\rm cm}^{-3}}\right) \right] \nonumber \\
& \times \left( \dfrac{r}{1\,{\rm kpc}}\right) \left(\dfrac{E}{10\mathrm{\,keV}}\right)^{-1}\, .
\label{eqa:qform}
\end{eqnarray}
Eq.~\eqref{eqa:prob} shows that for $\Phi\ll 1$, the dependence on the energy in the oscillation probability disappears, so that we recover again Eq.~(\ref{eq:penindep}).
Furthermore, in these conditions the probability grows quadratically with the distance, cancelling the dependence on the distance of the ALP flux. 
Of course, these results depend on the assumption that the magnetic field is correlated on a scale larger than $r$.

As obvious from Eq.~\eqref{eqa:qform}, for a massless ALP the condition $\Phi\ll 1$ is satisfied for distances $r \lesssim 20$~kpc at energies of $E\sim 10$~keV.
A finite ALP mass may spoil this condition. The exact mass threshold for this to happen depends on the distance $r$, and on the ALP energy.
In particular, high energy ALPs may be correlated on a larger distance and for larger masses. On the other hand, for large ALP masses $\Phi\gg 1$ and the oscillation probability acquires a dependence on the energy and decreases rapidly with the axion mass, $P_{a\gamma}\propto m_a^{-4}$, as evident in Figure~\ref{fig:prob}, where we show the behavior of the conversion probability $P_{a\gamma}$ for a source at distance $r=1$~kpc for a value of the regular magnetic field $B_{\rm reg}= 3$~$\mu$G in function of the ALP mass $m_a$.
In the Figure, we fixed the axion coupling to $g_{a\gamma}=10^{-11}\rm GeV^{-1}$ and take three representative ALP energies: $E=50$ keV (black curve), representative of the typical energy of ALPs emitted from a red supergiant star during its late evolutionary stages~\cite{Xiao:2020pra};  $E=100$ MeV (red curve), representative of the energies of ALPs emitted from supernova explosions~\cite{Payez:2014xsa}; and $E=10$ GeV (blue curve), representative of gamma-rays from galactic pulsars~\cite{Majumdar:2018sbv}.

\subsection{Conversions in the turbulent magnetic field}

A realistic description of the photon-ALP oscillation in the Galactic magnetic field requires a model for the magnetic field with both regular and turbulent components. 
The most commonly adopted model for the regular component is the  Jansson-Farrar model~\cite{Jansson:2012pc} (see, e.g., Refs.~\cite{Horns:2012kw,Galanti:2018upl, Majumdar:2018sbv}), though the Pshirkov model~\cite{Pshirkov:2011um} is a common alternative choice.

\begin{figure}[t!]
\vspace{0.cm}
\includegraphics[width=1\columnwidth]{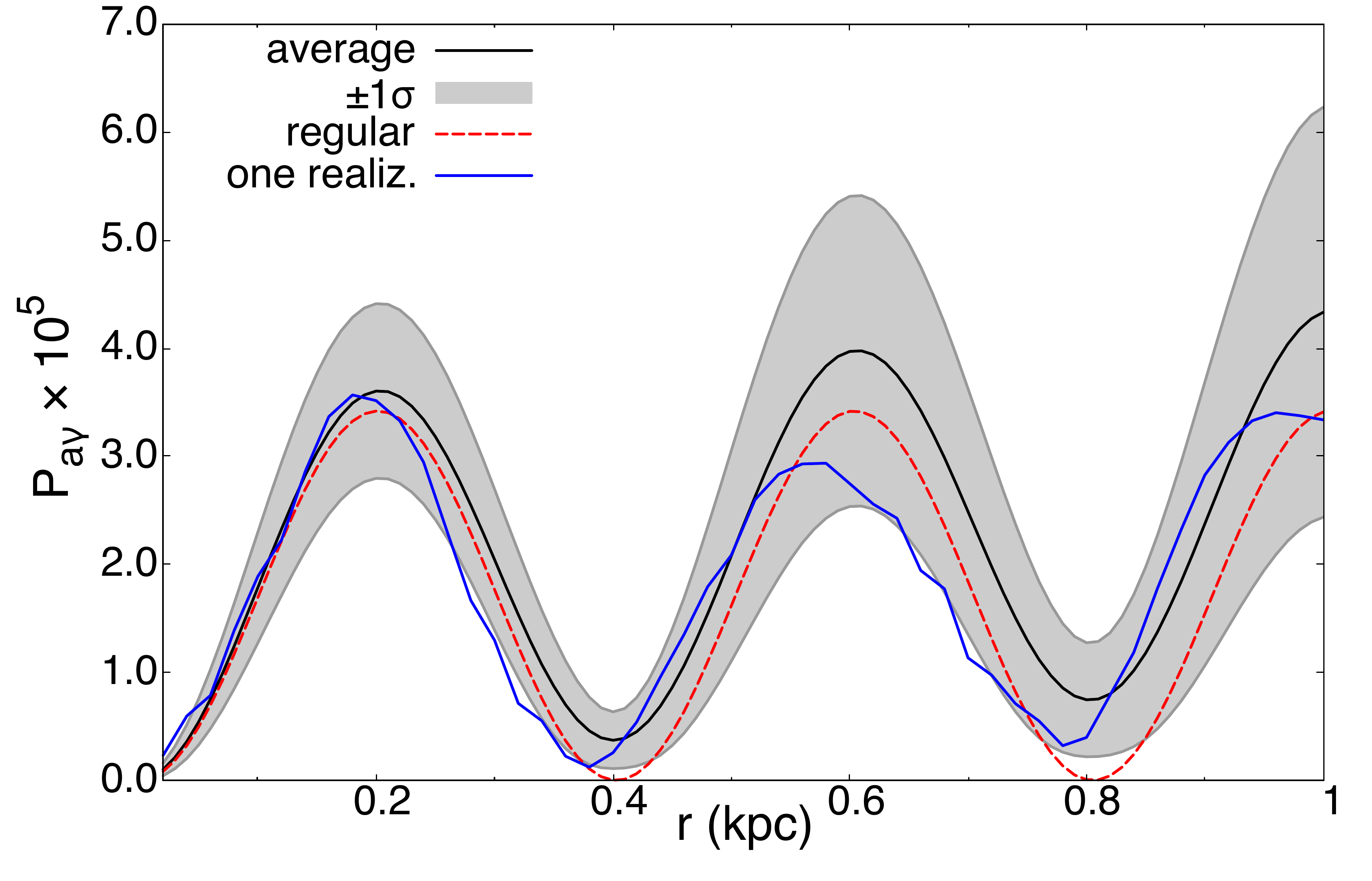}
\caption{ALP-to-photon conversion probability for  $g_{a\gamma}=10^{-11}$ GeV$^{-1}$, $m_a=10^{-10}$ eV and $E=50$ keV.
The dashed red curve corresponds to a pure regular $B$-field with $B_{\rm reg}=3$~$\mu$G. 
The black curve is the average probability for a turbulent field with $B_{\rm rms}=1$~$\mu$G and a correlation length $l_{\rm corr}=10$~pc, while the grey-band represents the ``$1\sigma$'' dispersion around the average. The blue line is the conversion probability for one possible random realization of the magnetic field. Note that $l_{\rm corr}=l_{\rm cell}/2$, where $l_{\rm cell}$ is the cell length.}
\label{fig:ma01neV}
\end{figure}

The small-scale turbulent field is less well known. It is usually assumed to follow a power law with a certain outer scale, where energy is injected, which then cascades down to smaller turbulent scales until energy dissipates at the dissipation scale~\cite{Evoli:2018nmb}. 
The strength of this random magnetic field component can be estimated from rotation measure fluctuations, combined with an estimate of the thermal electron density, giving a ratio of random to regular magnetic field  components of $\lesssim 1$~\cite{Haverkorn:2004kx,Schnitzeler:2007fc}. 
This value is compatible with the predictions of large-scale magnetic field models that include the turbulent magnetic field as a free parameter~\cite{Jaffe:2009hh}.

Concerning the correlation length, Ref.~\cite{Ohno1993} found a large $l_{\rm corr} \sim 100$~pc, using structure functions of rotation measures and averaging over large parts of the sky. Ref.~\cite{Haverkorn:2008tb} confirmed this large outer scale for interarm regions in the Galactic plane; however, they found a much smaller outer scale $\sim 10$~pc in the spiral arms. 
Also, arrival anisotropies in TeV cosmic-ray nuclei can be best explained by a turbulent magnetic interstellar medium on a maximum scale of about $\sim 1$~pc~\cite{Malkov:2010yq}. 
More recently, an upper limit of $\sim 20$~pc for the outer scale of the magnetic interstellar turbulence toward the Fan region has been found in Ref.~\cite{Iacobelli:2013fqa} by measuring with LOFAR the fluctuations in the diffuse synchrotron emission.

We now consider different representative situations showing the impact of the turbulent component of the Galactic magnetic field on the ALP-photon conversion. 
For definiteness, we take a source at a distance $r=1$~kpc. 
There are several  Galactic ALP sources, such as red supergiants, within a radius of ${\mathcal O}(1)$~kpc~\cite{Mukhopadhyay:2020ubs}.
ALPs produced in these sources would propagate in a magnetized medium where they can convert into photons.
In our examples below, we take the regular component of the magnetic field to be uniform on such a short length scales.
In any case, scales of 1 kpc are not resolved in large scales simulations. 
On the other hand, for the turbulent component we assume a correlation length $l_{\rm corr}=10$~pc.
The choice of a very short correlation length is conservative since, as we shall see below, increasing $l_{\rm corr}$ enhances the effect of the turbulent field on the conversion probability [Cf. Eq.~\eqref{eq:P1ave}].
In the description of the turbulent Galactic magnetic field we adopt a simplified approach, assuming it along a given line-of-sight as a network of domains with a fixed correlation length $l_{\rm corr}=10$~pc.~\footnote{A more refined characterization of turbulent $B$-fields in ALP conversions can be found in~\cite{Kartavtsev:2016doq}.}
Finally, we fix $B_{\rm reg}=3$ $\mu$G and assume a turbulent component with a root-mean-square (rms) amplitude $B_{\rm rms}=1$~$\mu$G~\cite{Jansson:2012pc}.
As a representative situation, we consider ALPs with coupling $g_{a\gamma}=10^{-11}$ GeV$^{-1}$, a value slightly below the current experimental and astrophysical  bounds, and energy $E=50$ keV. 
We will vary the ALP mass in order to probe different cases of ALP oscillation length. 
 
\begin{figure}[t!]
\vspace{0.cm}
\includegraphics[width=1\columnwidth]{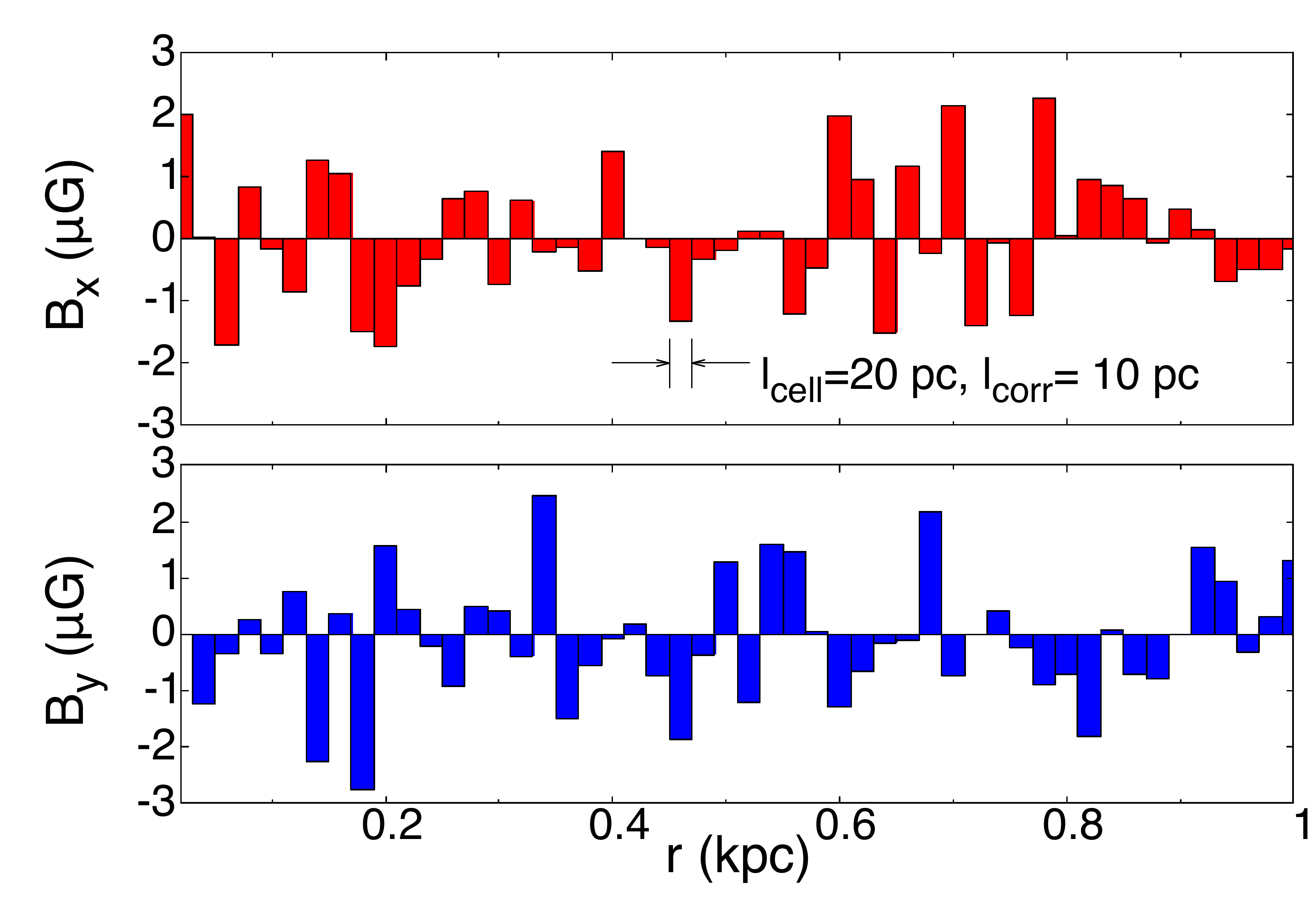}
\caption{Realization of the magnetic field used for the representative conversion probability with the blue line in Fig.~\ref{fig:ma01neV}.}
\label{fig:magnetic_realization}
\end{figure}

In the conditions described above, an ALP mass above about 1 neV corresponds to a correlation length  $l_{\rm osc} = 2 \pi/\Delta_{\rm osc} \simeq 4$~pc~$\ll l_{\rm corr}$.  
The associated conversion probability, in this case, is negligibly small, $P_{a\gamma} \lesssim 10^{-9}$. 
Therefore, we will ignore this case hereafter and limit ourselves to situations in which $l_{\rm osc} > l_{\rm corr}$. Within this hypothesis, we can show that  when the conversion probability is small, the effects  induced by the turbulent magnetic field can be evaluated analytically. As shown in Appendix A, the average probability over all possible field configurations in function of the distance $r$ from the source is remarkably simple, 
\begin{equation}
\langle P_{a\gamma}(r)\rangle=P_{a\gamma}^{(0)}(r)+g_{a\gamma}^2 B_{\rm rms}^2 l_{\rm corr} r\, ,
\label{eq:P1ave}\end{equation}
where $P_{a\gamma}^{(0)}(r)$ is the probability assuming only a regular magnetic field. We remark that Eq.~(\ref{eq:P1ave}) is valid also when the regular magnetic field is not homogeneous, as long as $P_{a\gamma}\ll 1$. 
This is the typical situation for conversions in the Milky Way, assuming allowed values for the ALP-photon coupling and a realistic magnetic field.
When $P_{a\gamma}\sim O(1)$ the previous approach is no longer valid and in general we have to resort to a Monte Carlo simulation. However, in the hypothesis of $\delta$--correlated perturbations it is possible to modify the evolution equations in order to calculate arbitrary moments of the variable $P_{a\gamma}$. 
Although this situation is outside the scope of this work, for completeness we treat this case in Appendix B.
 
A full derivation of Eq.~(\ref{eq:P1ave}) is provided in Appendix A.  However, it is easy to recognize the validity of this result also using very simple arguments.
Assuming $l_{\rm corr} \ll l_{\rm osc}$, it is possible to express the average conversion probability on a given magnetic domain as $P_{a\gamma, \rm turb}^{(0)}  \simeq (g_{a\gamma} B_{\rm rms} l_{\rm cell}/2)^2 = (g_{a\gamma}  B_{\rm rms} l_{\rm corr})^2 $, where the size of a single magnetic domain is given by $l_{\rm cell}=2 l_{\rm corr}$ (see Appendix A). Since the photons travel $N=r/l_{\rm corr}$ magnetic configurations, the incoherent total conversion probability on the turbulent configuration is given by  $P_{a\gamma, \rm turb} \simeq N P_{a\gamma, \rm turb}^{(0)} = g_{a\gamma}^2 B_{\rm rms}^2 l_{\rm corr} r$ (see also Ref.~\cite{Mirizzi:2007hr}).

We notice that since the turbulent field has a random configuration, $P_{a\gamma}$ is a random variable itself. 
In the hypothesis that the turbulent magnetic field has a gaussian distribution, $f(\widetilde{B}_{x,y})\propto \exp(-\widetilde{B}_{x,y}^2/2B_{\rm rms}^2)$, we can calculate the probability density function $F(P_{a\gamma})$ in the limit in which the diagonal terms in the Hamiltonian in Eq.~(\ref{eq:massgen}) are small (see Appendix A). 
If this condition is not fulfilled, it is very difficult to obtain a closed form for the probability distribution function (p.d.f.)  but it is still possible to get handy formulas for the calculation of the momenta of $P_{a\gamma}$. In particular one can obtain the second momentum of $P_{a\gamma}$ [Eq.~(\ref{eq:P2ave})] and thus the standard deviation of the distribution, $\sigma(r)=[\langle P^2_{a\gamma}(r)\rangle-\langle P_{a\gamma}(r)\rangle^2]^{1/2}$. 
In Appendix A we compare our analytical results with a numerical simulation to show the agreement between the two approaches. 

\begin{figure}[t!]
\vspace{0.cm}
\includegraphics[width=1\columnwidth]{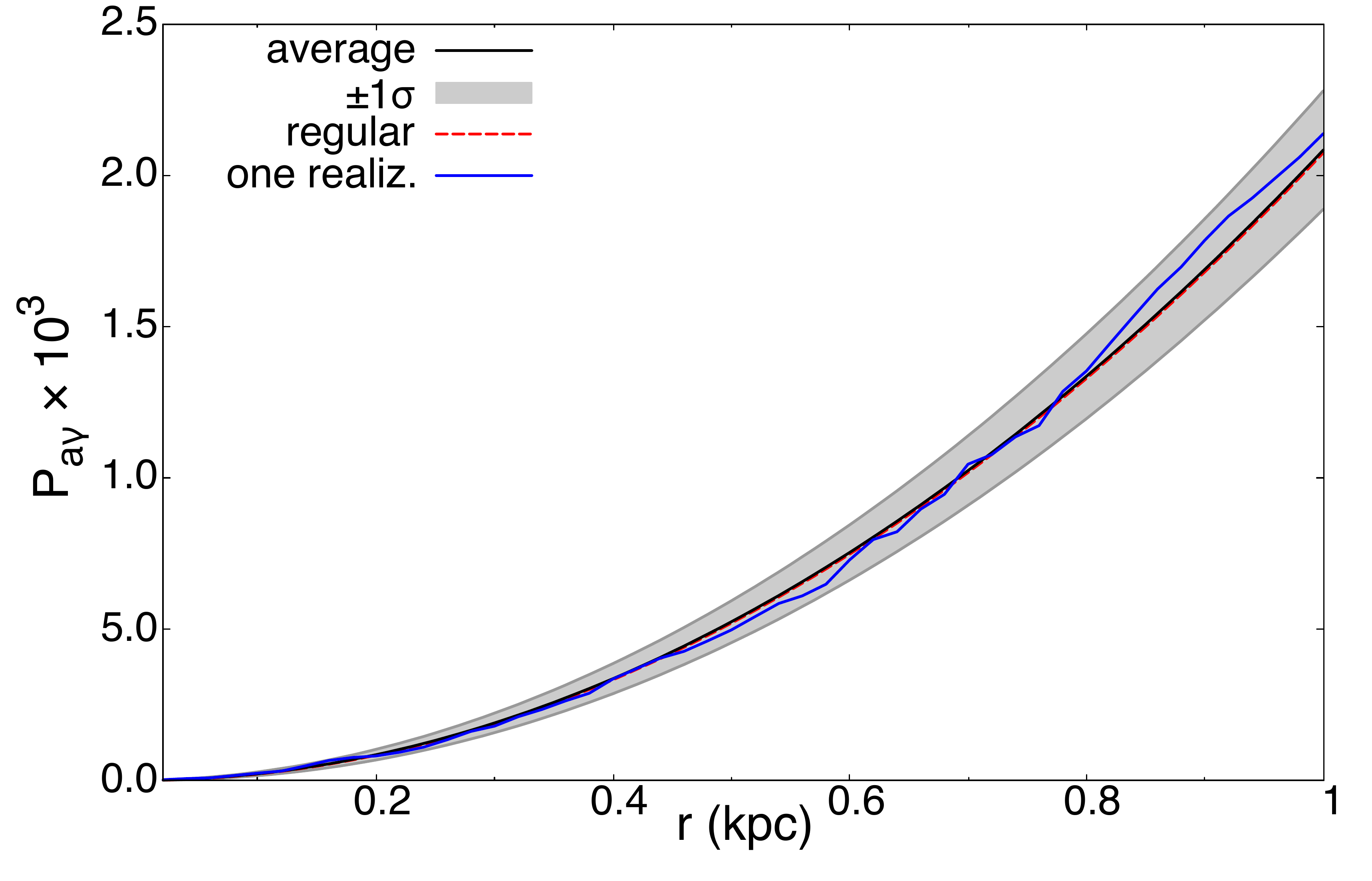}
\caption{ALP-to-photon conversion probability for $m_a=10^{-11}$~eV in the same format of Fig.~\ref{fig:ma01neV}.}
\label{fig:ma001neV}
\end{figure}

\begin{figure}[t!]
\vspace{0.cm}
\includegraphics[width=1\columnwidth]{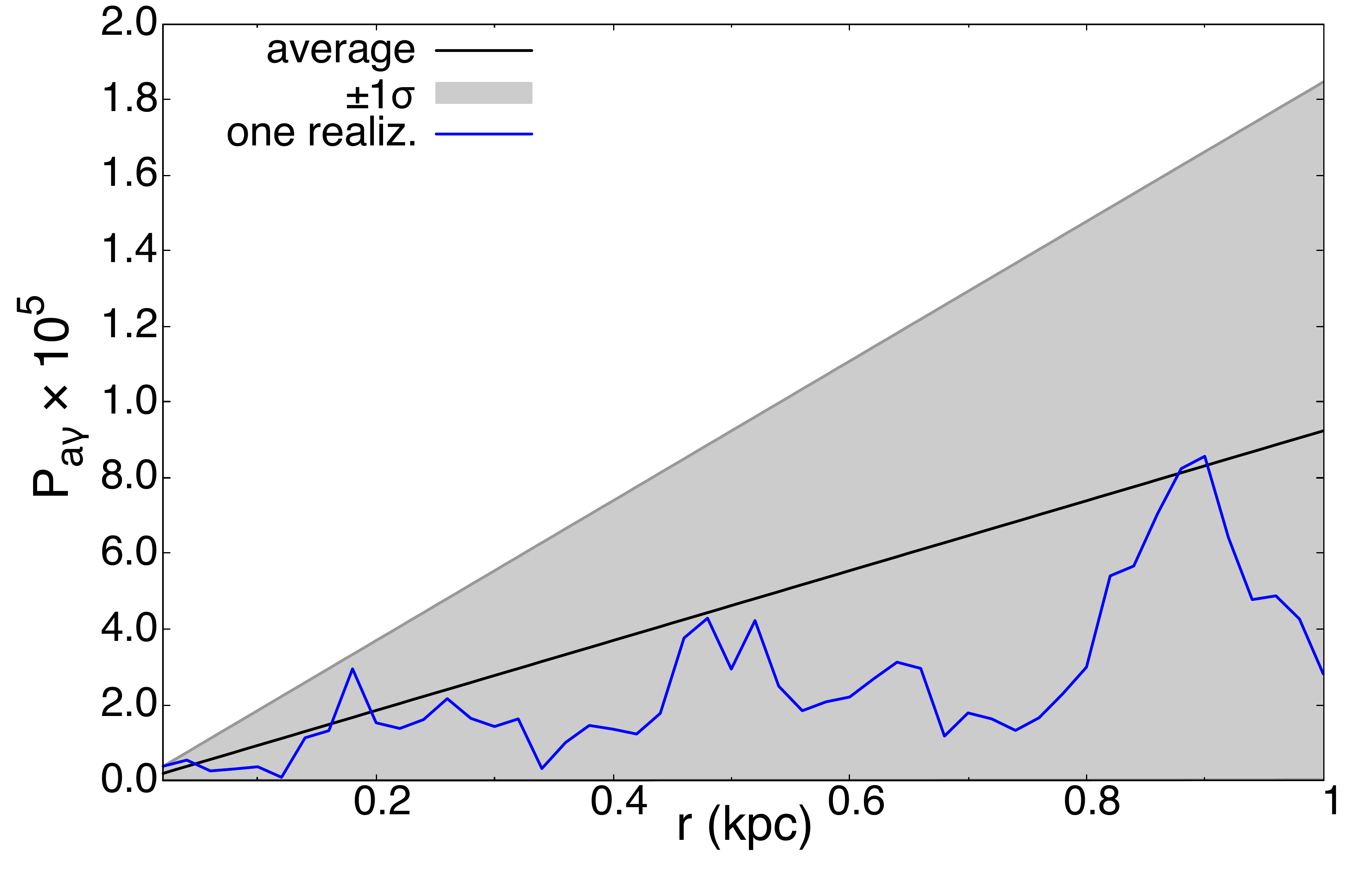}
\caption{ALP-to-photon conversion probability in the same format of  Fig. \ref{fig:ma001neV}, but with $B_{\rm reg}=0$.}
\label{fig:onlyturb}
\end{figure}

In Figs.~\ref{fig:ma01neV}, \ref{fig:ma001neV}, and \ref{fig:onlyturb}, we show examples of the ALP-photon oscillation probability in three different scenarios.
In particular, in Figure~\ref{fig:ma01neV} we consider the case $m_a=10^{-10}$~eV, corresponding to $l_{\rm osc} \simeq 400$~pc. The red curve is the conversion probability for a purely regular $B$-field. 
The black curve is the average probability in presence of regular plus turbulent fields, calculated using Eq.~(\ref{eq:P1ave}).
The gray band represents the ``$1\sigma$'' dispersion around the average represented by the grey-band.  
Here, by ``$1\sigma$'' we mean the standard deviation around the mean value. This cannot immediately translate into a definite confidence level since the p.d.f.\ is not in general a gaussian.
We see that the $1\sigma$ dispersion in the probability induced by the turbulent component is seizable, producing  variations up to  $\sim$50\%  with respect to the average.
For the purpose of illustration, with the blue line we show the conversion probability for one representative random realization of the magnetic field. The realization of the magnetic field used to draw this representative case is plotted in Fig.~\ref{fig:magnetic_realization}. We observe that the behavior of $P_{a\gamma}(r)$ for a single realization is in general irregular, depending on the configuration of the field. In particular, the chosen realization for $r\gtrsim 0.5$~kpc gives a conversion probability smaller than the purely regular field. In general, some realizations of the turbulent field could result in a conversion probability smaller than the one obtained with a purely regular field. However, the average probability is always larger compared to a regular field.

In Figure \ref{fig:ma001neV}, we consider an ALP with mass $m_a=10^{-11}$ eV, which implies $l_{\rm osc} \simeq 35$~kpc. 
This case corresponds to an energy-independent oscillation probability. As expected, the conversion probability induced by the regular component of the magnetic field scales as $r^2$. The effect of the turbulent component is the spread of the $P_{a\gamma}$ around the mean. 
Although on average the effect is very small, for many realizations the difference with the regular case can exceed up to 10\% within $1\sigma$. Since in this case $\Delta_{\rm pl}, \Delta_a\ll \Delta_{a\gamma}$ we can use the analytic expression for the p.d.f.\ [Eq.~(\ref{eq:PDF}) in Appendix A]. For example, we estimate that at distance $r=1$~kpc about the 31.7\% of the values of $P_{a\gamma}$ lie outside the $\pm1\sigma$ band. 

Finally, in order to isolate the role of the turbulent field, in Fig.~\ref{fig:onlyturb} we consider the same case  of  Fig.~\ref{fig:ma001neV}, but  assuming  only the turbulent component of the magnetic field.  
We show the average and the $1\sigma$ band for a pure turbulent field with  $B_{\rm rms}=1$~$\mu$G. In this case, as expected from Eq.~(\ref{eq:P1ave}),
we see that $\langle P_{a\gamma}(r)\rangle$ grows linearly (at least until $\langle P_{a\gamma}\rangle\ll 1$). Also in the last two cases in blue we draw a representative conversion probability curves, using the same realizations drawn in Fig.~\ref{fig:magnetic_realization}.

\section{Effect on observable photon spectra}
\label{sec:Effect_photon_spectra}

In this Section, we assess the impact of the turbulent component of the Galactic magnetic field in three physically relevant examples.
Specifically, we consider \emph{i)} the case of ALPs produced in the red supergiant Betelgeuse, which can oscillate into photons originating a hard X-ray flux; \emph{ii)} ALPs produced in a galactic supernova (SN), which can originate a gamma-ray flux; and \emph{iii)} ALPs produced by conversions from gamma-rays from Galactic pulsars.
All these cases have already been considered in the literature. However, the impact of the turbulent magnetic field has always been ignored. 

The signatures of the random component of the magnetic field can be revealed through energy-dependent irregularities imprinted on the photon energy spectra. As we shall see, these effects can be more or less pronounced,  depending on the ALP parameters, particularly the ALP mass.   
In general, we expect the impact of the turbulent field to be maximal at the threshold between the energy-dependent and the energy-independent regime [see Eq.~(\ref{eqa:prob})], where the conversion probability reaches its maximum before saturating (see Fig. 1).

In all of the examples in this section, we will assume a turbulent field with a strength $B_{\rm rms}= 1$ $\mu$G and correlation length $l_{\rm corr}= 10$~pc. For simplicity we will consider a single realization of the turbulent magnetic field, different for the three physical cases, avoiding to characterize a distribution obtained generating different configurations of the field. In all the examples the regular magnetic field is parametrized according to the Jansson and Farrar model~\cite{Jansson:2012pc} along different lines-of-sight.

 \begin{table}[!t]
\begin{center}
\begin{tabular}{lccc}
\hline
 & 
$C$
& $E_0$ [keV] &$\beta$ \\
\hline
\hline
Betelgeuse& $1.36    $ & $50$ &1.95 \\
 SN& $1.75 \times 10^{5} $ & $1.20\times10^{5}$ &2.40 \\
\hline
\end{tabular}
 \caption{Fitting parameters for the Betelgeuse and  SN ALP spectrum from the Primakoff process.
 }
\label{tab:fitting}
\end{center}
\end{table}

To start, we consider Betelgeuse, a red supergiant star of about $20\,M_\odot$, at a distance of $200$~pc. Betelgeuse's exact evolutionary phase is unknown, except for the fact that it has already exhausted the H in its core. Here, we assume that the star is in the core He-burning stage, which is the most likely scenario since the following stages are considerably shorter. For the numerical model, we refer to~\cite{Xiao:2020pra}. ALPs can be thermally produced in Betelgeuse through the Primakoff process.
Their production rate can be expressed as a quasi-thermal spectrum
\begin{equation}
\frac{d\dot{N}_{a}}{dE} = \frac{10^{42}Cg_{11}^{2}}{\textrm{keV}~\textrm{s}}\left(\frac{E}{E_{0}}\right)^{\beta}e^{-(\beta +1)E/E_{0}} \,\ ,
\label{eq:spectrum}
\end{equation}
where $C$ is a normalization factor, $E_0$ represents the average energy, and $\beta$ is the spectrum index. 
The values of $C$, $E_0$ and $\beta$ depend on various structural parameters characterizing the core of the star, such as temperature, density and chemical composition. Here, we adopt the parameters in Table~\ref{tab:fitting}, corresponding to Model 0 in Table S1 of~\cite{Xiao:2020pra} 

After being produced, the ALPs leave the star unimpeded and may convert into photons in the Galactic magnetic field. From the average ALP energy in Table~\ref{tab:fitting} it is clear that the resulting photons are expected to have a hard X-ray spectrum. Such a spectrum was searched, with negative results, in a recent dedicated NuSTAR observation~\cite{Xiao:2020pra}, 
leading to the bound 
$g_{a\gamma} < 2 \times 10^{-11}$ GeV$^{-1}$ for $m_a < 5 \times 10^{-11}$ eV. The analysis, however, ignored completely the random component of the magnetic field. 
 
\begin{figure}[t!]
\vspace{0.cm}
\includegraphics[width=0.95\columnwidth]{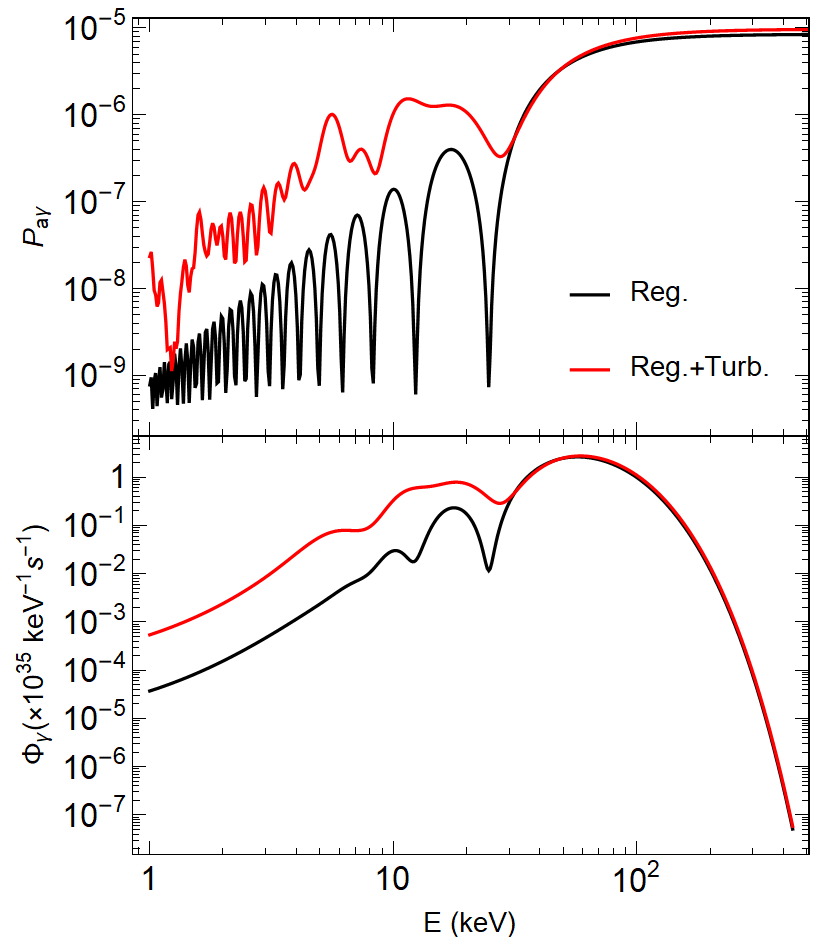}
\caption{
Upper panel: ALP-photon conversion probability for the red supergiant star Betelgeuse at $r=200$~pc in function of the energy for  $g_{a\gamma}=10^{-11}$ GeV$^{-1}$ and $m_a=10^{-10}$ eV for regular (black curve) and regular plus turbulent (red curve) magnetic field. 
Lower panel: Photon energy spectrum after ALP conversions in  regular (black curve) and regular plus turbulent (red curve) magnetic field. A gaussian energy resolution of 1~keV is assumed.
}
\label{fig:HB spectrum}
\end{figure}

In the upper panel of Figure~\ref{fig:HB spectrum}, we show the ALP-photon conversion probability vs.\ the energy, for  $g_{a\gamma}=10^{-11}$ GeV$^{-1}$ and $m_a=10^{-10}$ eV.
The black curve includes only the regular magnetic field. In this case, we notice an almost periodic behavior for $E\lesssim 30$ keV, while at higher energies the energy-independent plateau is reached. 
The effects of the turbulent component are shown with the red curve in the same figure. In this case, we notice not only a larger conversion probability, expected since the total magnetic field strength is (in average) enhanced. We also see that the periodic pattern observed in the black curve is destroyed and replaced with an irregular behavior.

In the lower panel of Figure \ref{fig:HB spectrum}, we show the observable photon number flux, obtained by convolving the original ALP flux with the $P_{a\gamma}$,
\begin{equation}
\Phi_{\gamma}(E) \equiv \frac{d N_\gamma}{d E} = \frac{d\dot{N}_{a}}{dE}  \times P_{a\gamma} \,\ .
\end{equation}
We also consider a gaussian energy resolution of 1~keV, comparable with the performance of NuSTAR~\cite{nustar}. We observe that the irregular behavior is smeared out by the resolution effect. We find peculiar features in the energy range [10:40] keV, appearing as peculiar bumps and dips. 
Thus, in principle, the case of just regular field is distinguishable from the case of regular plus turbulent fields.

As a second example, we consider the ALP flux produced by Primakoff process in supernovae. 
The ALP production rate is calculated as in Ref.~\cite{Calore:2020tjw}, using a SN model with a 18 $M_{\odot}$ progenitor, simulated in spherical symmetry with the AGILE-BOLTZTRAN code~\cite{Mezzacappa:1993gn,Liebendoerfer:2002xn}. 
We consider the rate at $t_{pb}=1$~s after the bounce. Even in this case, the ALP production rate has the quasi-thermal spectrum given in Eq.~(\ref{eq:spectrum}). The fitting parameters corresponding to the model we are considering are also given in Table~\ref{tab:fitting}.

Because of the conversion in the Galactic magnetic field, the ALP flux generates a gamma ray flux, with typical energies of O(100) MeV.  
The lack of a gamma-ray signal in the Gamma-Ray Spectrometer (GRS) on the Solar Maximum Mission (SMM) in coincidence with the observation of the neutrinos emitted from SN 1987A therefore provided a bound on ALPs coupling to photons~\cite{Brockway:1996yr,Grifols:1996id}. 
Specifically, for $m_a=4 \times10^{-10}$ eV the most recent analysis finds $g_{a\gamma}= 5.3 \times 10^{-12}$ GeV$^{-1}$~\cite{Payez:2014xsa}. 
A future ALP burst from a Galactic SN would allow to probe a considerable larger parameter space through a Fermi-LAT observation, if the explosion occurs in its field of view~\cite{Meyer:2016wrm}.

\begin{figure}[t!]
\vspace{0.cm}
\includegraphics[width=0.95\columnwidth]{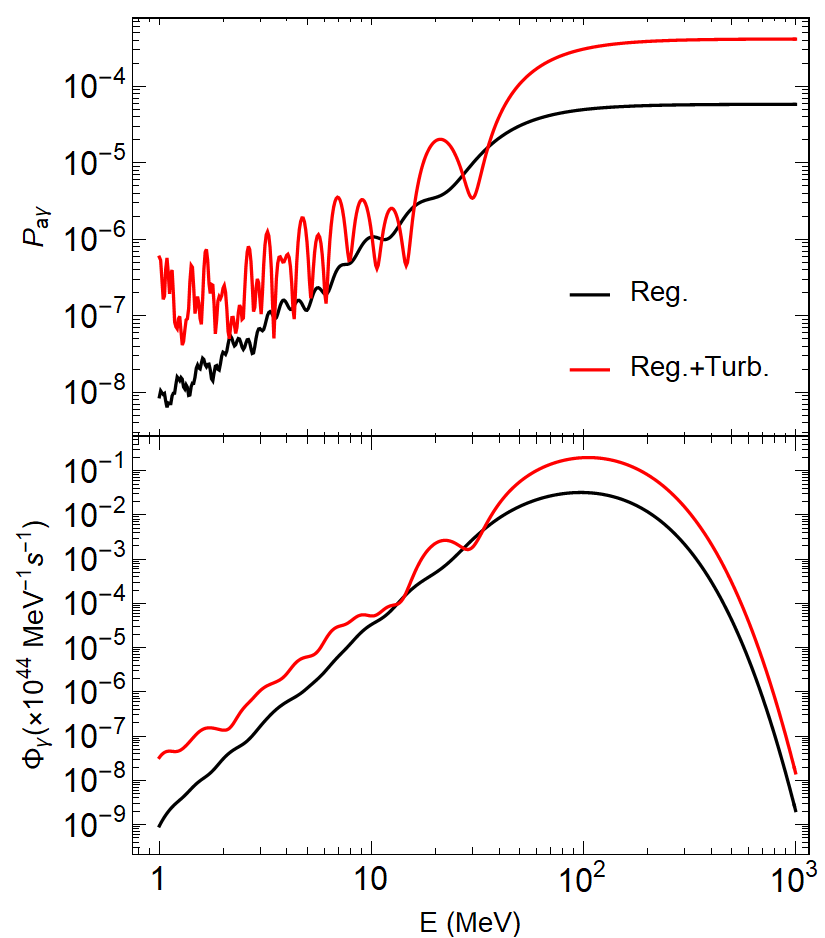}
\caption{Upper panel: ALP-photon conversion probability for  a Galactic supernova at $r=10$~kpc in function of the energy for  $g_{a\gamma}=10^{-12}$ GeV$^{-1}$ and $m_a=5 \times10^{-10}$ eV for regular (black curve) and regular plus turbulent (red curve) magnetic field. 
Lower panel: Photon energy spectrum after ALP conversions in  regular (black curve) and regular plus turbulent (red curve) magnetic field. 
A gaussian energy resolution of 10$\%$ is assumed.}
\label{fig:SN spectrum}
\end{figure}

Here we want to assess the impact of the turbulent magnetic field of these potential observations. 
For definitiveness, we assume a Galactic SN at $r=10$~kpc. We consider a specific line of sight in the Milky Way $(b,l)=(75.22^{\circ}, 0.11^{\circ})$. 
In the upper panel of Figure \ref{fig:SN spectrum} we show the ALP-photon conversion probability in function of the energy for  $g_{a\gamma}=10^{-12}$ GeV$^{-1}$ and $m_a=5 \times10^{-10}$ eV for regular (black curve) and regular plus turbulent (red curve) magnetic field. 
We observe peculiar wiggles in the energy range $E\in [1:10^2]$~MeV. Larger irregularities are observed in the presence of the turbulent field. In the lower panel we present the corresponding photon spectra assuming an energy resolution of 10$\%$, as expected for Fermi-LAT at those energies~\cite{fermi}. We realize that in the presence of only regular field the energy-dependent modulation is almost washed out by the effect of the resolution, while in the presence of the turbulent component bumpy features below $E\sim 100$ MeV would be clearly visible. We notice that this energy range would be below Fermi-LAT sensitivity, but is in the reach of future gamma-ray experiments like eASTROGAM~\cite{DeAngelis:2017gra} and AMEGO~\cite{McEnery:2019tcm}.

\begin{figure}[t!]
\vspace{0.cm}
\includegraphics[width=0.95\columnwidth]{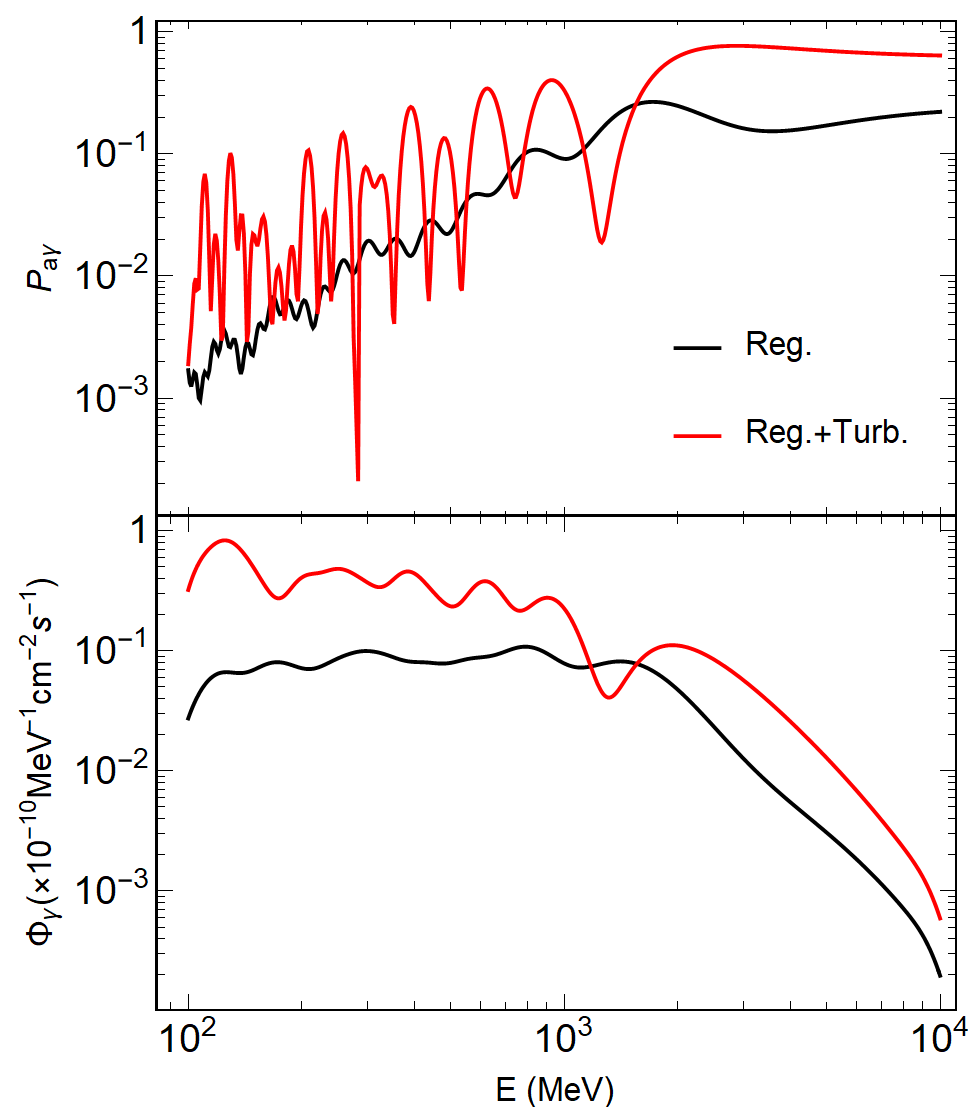}
\caption{Upper panel: Photon-ALP conversion probability for the  Galactic pulsar PSR J2021+3651 at distance $r=6$~kpc for  $g_{a\gamma}=3.5 \times 10^{-12}$ GeV$^{-1}$ and $m_a=4.4 \times10^{-9}$ eV for regular (black curve) and regular plus turbulent (red curve) magnetic field. 
Lower panel: Photon energy spectrum after ALP conversions in  regular (black curve) and regular plus turbulent (red curve) magnetic field. 
A gaussian energy resolution of 10$\%$ is assumed.}
\label{fig:PSR spectrum}
\end{figure}

Finally we consider the case of Galactic pulsars. This case is particularly significant since it was recently claimed by two different groups~\cite{Majumdar:2018sbv,Xia:2018xbt}  that the unexpected spectral  modulation observed by Fermi-LAT in gamma-rays from Galactic pulsars and supernova remnants could be due to conversion of photons into ultra-light ALPs in the Milky-Way magnetic field. 
The best-fit ALP parameters,  $g_{a\gamma}= 2.4 \times 10^{-10}$ GeV$^{-1}$  and $m_a \sim 4  \times10^{-9}$ eV, are in tension with the CAST bound on solar ALPs.
The tension can be lifted in ALP models with environmental dependent mass/coupling~\cite{Pallathadka:2020vwu}. Here, we examine the effects of the previously neglected turbulent magnetic field on these results. For definitiveness, we consider the pulsar PSR J2021+3651 at distance $r=6$~kpc from us and with Galactic coordinates $ (l,b) = (75.22, 0.11)$ studied in Ref.~\cite{Majumdar:2018sbv}. The original photon spectrum is given by
\begin{equation}
\frac{dN^0_{\gamma}}{d E} = N_0
 \left(\frac{E}{E_0} \right)^{-\Gamma} \exp\left(-\frac{E}{E_{\rm cut}} \right) \,\ ,
\end{equation}
where $N_0= 0.15 \times 10^{-9}$~{MeV}$^{-1}$~{cm}$^{-2}$~{s}$^{-1}$, $E_0=0.8$~GeV, $\Gamma=1.59$, $E_{\rm cut}=3.2$~GeV.
In the upper panel of Figure \ref{fig:PSR spectrum}, we show the photon-ALP conversion probability for PSR J2021+3651 for  $g_{a\gamma}=3.5 \times 10^{-12}$ GeV$^{-1}$ and $m_a=4.4 \times10^{-9}$ eV corresponding to the best-fit for this source~\cite{Majumdar:2018sbv} for regular (black curve) and regular plus turbulent (red curve) magnetic field. We recognize that energy modulations are visible in the range $E\in[10^2:10^3]$ MeV, and strongly enhanced when a turbulent component is also present. 
In the lower panel, we show the photon spectrum after photon-ALP conversions, i.e.
 \begin{equation}
 \Phi_\gamma(E) \equiv \frac{dN_{\gamma}}{d E} =\frac{dN^0_{\gamma}}{d E} \times (1-P_{a\gamma}) \,\ .
 \end{equation}
%
A gaussian energy resolution of 10$\%$ is assumed. It is very clear how the modulations, already visible in the case of regular field, are magnified by the presence of the turbulent component. This result would suggest that the analysis of~\cite{Majumdar:2018sbv} might deserve a revisitation with the inclusion of the turbulent component of the Galactic $B$-field.

\section{Conclusions}
\label{sec:Conclusions}

The magnetic field in the Milky Way has been recognized as a very powerful tool to probe the low mass ALP parameter space, since it leads peculiar ALP-induced signatures in observable photon spectra from different (extra)galactic sources. 
The current literature has adopted state-of-the-art models to characterize the morphology of the regular component of Galactic magnetic field. 
However, the small-scale turbulent component has been neglected so far, in spite of having a strength comparable to the regular one.
In our work, we have investigated this important aspect providing numerical and analytical recipes to characterize the impact of the turbulent component of the Galactic magnetic field on the ALP-photon conversions. Here, we summarize and discuss our main results. First, we have shown that on average the effect of the turbulent magnetic field on photon-ALP conversions grows linearly with the distance from the source at least until the conversion probability remains small. 
This result is summarized in Eq.~\eqref{eq:P1ave}. Furthermore, we have shown that the effect of the turbulent component are especially relevant around the transition energy between the energy-dependent and the energy-independent regime, where the conversion probability reaches its maximum before saturating. Intriguingly, this transition energy selects a specific range of the ALP mass for which we expect energy-dependent irregularities to be imprinted on the photon energy spectra.
We have shown that these irregularities could  be observable in the photon spectra associated to different galactic sources, such as red supergiants, supernovae and pulsar, on an energy range between 100 keV and 100 GeV.
If such  features were to be detected in a future observation, that would represent a direct signature of the turbulent component of the Galactic $B$-fields. 
Moreover, it would allow to constraint the ALP mass range, which is typically a very difficult task. Our study might be relevant also in other contexts, where one expects a combination of regular and turbulent magnetic fields.
An example is the case of Galaxy Clusters, as recently discussed in Ref.~\cite{Libanov:2019fzq}. 

In conclusion, our results show that neglecting the random component of the Galactic magnetic field is not always justified, even in cases when its correlation length is much smaller than the ALP oscillation length. Moreover, we have shown how the random component leads to recognizable features in the photon spectrum, which could reveal information about the ALP parameters and the magnetic field itself. All these results confirm once more the high physics potential in gamma-ray observations to constrain the ALP parameter space.

\acknowledgments

We warmly thank Francesca Calore for reading the manuscript and for useful comments on it.
The work of P.C., A.M. and D.M.  is partly supported by the Italian Ministero dell’Universit\`a e Ricerca (MUR) through the research grant no. 2017W4HA7S ``NAT-NET: Neutrino and Astroparticle Theory Network'' under the program PRIN 2017, and by the Istituto Nazionale di Fisica Nucleare (INFN) through the ``Theoretical Astroparticle Physics'' (TAsP) project.

\appendix

\section{Gaussian turbulent component--perturbative approximation}

We can decompose the transverse ${\bf B}_T$ field in two components, a regular one and a turbulent one: ${\bf B}_T={\bf B}_{\rm reg}(z)+\widetilde{\bf B}(z)$, with $\langle \widetilde{\bf B}(z)\rangle=0$ where $\langle ...\rangle$ denotes an average on all possible field configurations. We suppose that the two transverse components of the turbulent field are totally uncorrelated, $\langle \widetilde{B}_x(z_1)\widetilde{B}_y(z_2)\rangle=0$, while we can define the two-ponits correlation function $\langle \widetilde{B}_x(z_1)\widetilde{B}_x(z_2)\rangle\equiv C_B(z_1-z_2)$. Reasonably, the correlation functions for the $x$ and $y$ components are identical. The correlation function is in general a function that goes rapidly to zero for $|z_1-z_2|\gtrsim l_{\rm corr}$. We can conveniently define the {\it correlation length} $l_{\rm corr}$ as
\begin{equation}
l_{\rm corr}=\frac{1}{B_{\rm rms}^2}\int_0^\infty d\xi\, C_B(\xi)\, ,
\label{eq:corrlen}
\end{equation}
where $B_{\rm rms}^2=C_B(0)$ is the variance of the turbulent magnetic field. 

The simplest model of turbulent field is the ``cell model'' in which the space is divided into cells with dimension $\sim l_{\rm cell}$. In each cell $\widetilde{\bf B}$ has a constant random value with zero mean and variance $B_{\rm rms}$ and uncorrelated to the value of the field in adjacent cells. In this case is easy to show that $C_B(\xi)=B_{\rm rms}^2 \cdot (1-|\xi|/l_{\rm cell})$ for $|\xi|\leq l_{\rm cell}$ and zero otherwise. In fact, if $|\xi|\geq l_{\rm cell}$ the points $z$ and $z+\xi$ always fall in different cells and thus their correlation must vanish. Conversely, when $|\xi|<l_{\rm cell}$, assuming that the point $z$ falls in a given cell, the correlation is proportional to the probability that the point $z+\xi$ falls in the same cell. Is easy to realize that this probability is just the ratio $(l_{\rm cell}-|\xi|)/l_{\rm cell}$. 

Using the definition in Eq.~(\ref{eq:corrlen}) we see that the correlation length is half of the cell length, $l_{\rm corr}=l_{\rm cell}/2$. Although this model is very handy for practical purposes, it is unrealistic since cannot satisfy the condition $\nabla\cdot {\bf B}=0$ on the boundary of cells. 

A more realistic model is a Kolmogorov-like power-law spectrum whose three-dimensional Fourier transform of the correlation function is given by
\begin{equation}
\hat C_{B,ij}({\mathbf q}-{\mathbf q}')=(2\pi)^6 M(|{\mathbf q}|)\cdot\left(\delta_{ij}-\frac{q_i q_j}{|{\mathbf q}|^2}\right)\delta^3({\mathbf q}-{\mathbf q}')\, ,
\end{equation}
with $M(|{\mathbf q}|)\sim |{\mathbf q}|^{-\alpha}$. The one dimensional correlation function can be obtained as in~\cite{Meyer:2014epa} [Eqs.~(A.9--16)].

For the sake of illustration we tale the regular component constant and parallel to the $y$-axis without loss of generality, although the results can be easily extended to the case of non constant ${\bf B}_{\rm reg}$. We notice also that the conversion probability $P_{a\gamma}$ is generally very small in the Galaxy due to the relatively short distances travelled by the photon. Within this approximation we can solve Eq.~(\ref{we}) perturbatively. 

\begin{figure*}[t!]
\vspace{0.cm}
\includegraphics[width=2\columnwidth]{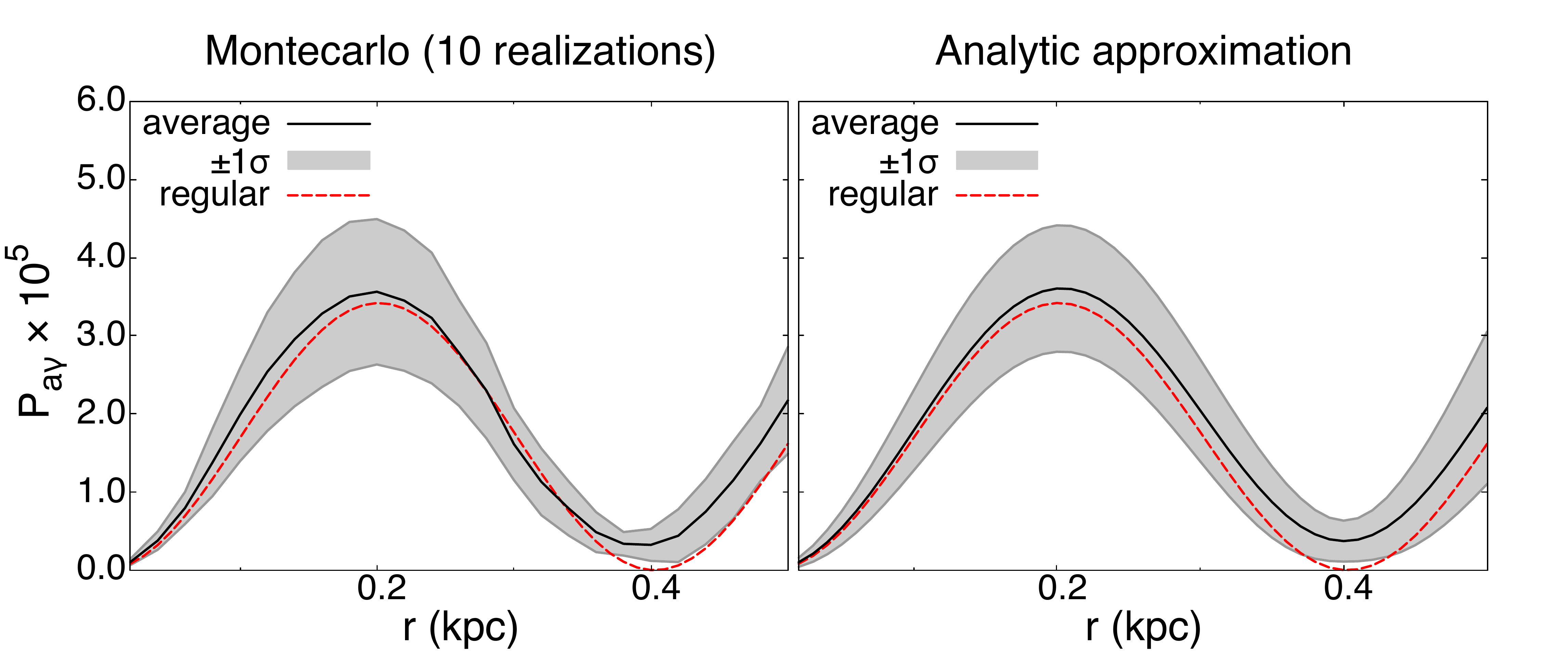}
\caption{Left plot: Average conversion probability and $\pm1\sigma$ band like Figure \ref{fig:ma01neV} obtained through a Monte Carlo (10 realizations). Right plot: same as left but using Eqs.~(\ref{eq:Pave}), (\ref{eq:lin}) and (\ref{eq:P2ave}).
}
\label{fig:comparison}
\end{figure*}

We start from the solution of Eq.~(\ref{we}) for $A_{x,y}(z)$ 
\begin{equation}
A_{x,y}(z)=-ie^{-ik z}\int_0^z d\zeta\, e^{ik\zeta}\Delta_{x,y}(\zeta) a(\zeta)\, ,
\end{equation}
with $\Delta_x=\frac{1}{2}g_{a\gamma}\widetilde{B}_x$, $\Delta_y=\frac{1}{2}g_{a\gamma}(B_{\rm reg}+\widetilde{B}_y)$,  and $k=\Delta_{\rm pl}+\Delta_{\rm QED}-\Delta_a$. For an initial ALP state we can neglect the back-conversion of photons into ALPs, $a(z)\simeq a(0)=1$. 
The ALP to photon conversion probability can be approximatively written as
\begin{eqnarray}
P_{a\gamma}(z)&=&|A_x(z)|^2+|A_y(z)|^2\nonumber \\
&=&P_{a\gamma}^{(0)}+|{\cal T}_x(z)|^2+|{\cal T}_y(z)|^2+2\sqrt{P_{a\gamma}^{(0)}}{\rm Re}[{\cal T}_y(z)]\, ,\nonumber\\
\label{eq:Pagamma}
\end{eqnarray}
with
\begin{equation}
P_{a\gamma}^{(0)}(z)=\left(\frac{g_{a\gamma}B_{\rm reg}}{k}\right)^2\,\sin^2\frac{kz}{2}\, ,
\end{equation}
and
\begin{equation}
{\cal T}_{x,y}(z)=\int_0^z d\zeta\,\frac{1}{2}g_{a\gamma}\widetilde{B}_{x,y}(\zeta)e^{ik\left(\zeta-\frac{z}{2}\right)}\, .
\end{equation}
Averaging on all possible configurations of the turbulent field we have 
\begin{equation}
\langle P_{a\gamma}(z)\rangle=P_{a\gamma}^{(0)}(z)+\Delta P(z)\, ,
\label{eq:Pave}\end{equation}
with 
\begin{eqnarray}
\Delta P(z)&=&\frac{1}{4}g_{a\gamma}^2\int_0^z d\zeta_1 d\zeta_2\, \langle\widetilde{B}_x(z_1)\widetilde{B}_x(z_2)\rangle e^{ik(\zeta_1-\zeta_2)}\nonumber\\
&=&g_{a\gamma}^2\int_0^z d\xi (z-\xi)\cdot C_B(\xi)\cdot\cos k\xi\, ,
\end{eqnarray}
where we have used the Cauchy formula for nested integrals. For $z\gg l_{\rm corr}$ the upper limit in the integral can be approximated to infinity and thus $\Delta P(z)$ has a simple linear form, $\Delta P(z)\simeq A+Bz$.

A remarkable simplification can be obtained when the correlation length is much smaller than the oscillation wavelength $l_{\rm corr}\ll k^{-1}$. In this case we can approximate the correlation function to a $\delta$ function, $C_B(\xi)\simeq 2B_{\rm rms}^2 l_{\rm corr}\delta(\xi)$, and thus 
\begin{equation}
\Delta P(z)\simeq g_{a\gamma}^2 B_{\rm rms}^2 l_{\rm corr} z\, ,
\label{eq:lin}\end{equation}
valid when $\Delta P(z)\ll 1$ (see also \cite{Mirizzi:2007hr} for a similar derivation.

The second momentum of the distribution can be calculated by squaring Eq.~(\ref{eq:Pagamma}) and taking the average. Further simplifications can be obtained in the hypothesis that the turbulent component has a gaussian distribution. In this case it can be shown that the $n$-points correlators are vanishing for $n$ odd and are the sum of all permutations of product of 2-point correlators for $n$ even, e.g.
\begin{eqnarray}
\langle \widetilde{B}_x^{(1)}\widetilde{B}_x^{(2)}\widetilde{B}_x^{(3)}\widetilde{B}_x^{(4)}\rangle
&=&C_B(z_1-z_2)\,C_B(z_3-z_4)\nonumber\\
&+&C_B(z_1-z_3)\,C_B(z_2-z_4)\nonumber\\
&+&C_B(z_1-z_4)\,C_B(z_2-z_3)\, , \hspace{0.8cm}\,
\label{eq:deltacorr2}
\end{eqnarray}
where $\widetilde{B}_x^{(k)}\equiv \widetilde{B}_x(z_k)$. For gaussian $\delta$-correlations, after straightforward calculations we have
\begin{eqnarray}
\langle P^2_{a\gamma}(z)\rangle=P_{a\gamma}^{(0)^2}&+&P_{a\gamma}^{(0)} \Delta P\left[3+\sinc(kz)\right]\nonumber\\
&+&\frac{1}{2} \Delta P^2\left[3+\sinc^2(kz)\right]\, ,
\label{eq:P2ave}
\end{eqnarray}
where $\Delta P$ is given by Eq.~(\ref{eq:lin}). From this relation we can obtain the standard deviation of the distribution, $\sigma(z)=\sqrt{\langle P^2_{a\gamma}(z)\rangle-\langle P_{a\gamma}(z)\rangle^2}$.

In Figure \ref{fig:comparison} we compare the results obtained through a Monte Carlo simulation with 10 realizations of the magnetic field with the analytic formulas in Eqs.~(\ref{eq:Pave}), (\ref{eq:lin}) and (\ref{eq:P2ave}). For the Monte Carlo simulation we consider a cell-like structure with $l_{\rm cell}=20$~pc. The regular field is chosen $B_{\rm reg}=3$~$\mu$G while for each cell the transverse magnetic field has a random direction and a random strength with gaussian distribution and a r.m.s 
$B_{\rm rms}=1$~$\mu$G~\footnote{A practical way to obtain two random generated components with gaussian distribution with $0$ mean and $B_{\rm rms}^2$ variance is to use the Box-Muller theorem:
$B_{T,x}=B_{\rm rms}\sqrt{-2\log U}\cos(2\pi V)$, $B_{T,y}=B_{\rm rms}\sqrt{-2\log U}\sin(2\pi V)$ with $U$, $V$ are random numbers with uniform distribution in the interval $]0;1[$~\cite{Box}.}, 
like in Figure \ref{fig:ma01neV}. Since the oscillation length for the regular field is $l_{\rm osc}\simeq 400$~pc ($\gg l_{\rm cell}$), we can safely consider the turbulent component as $\delta$--correlated. In the right plot we calculate the average probability and $1\sigma$ dispersion using Eqs.~(\ref{eq:Pave}), (\ref{eq:lin}) and (\ref{eq:P2ave}). The grey band represents the $\pm 1\sigma$ spread around the average. We remark that this band does not represent a definite confidence level since the distribution is in general not gaussian, as we will see soon. We notice the good agreement between the numerical simulation and the analytic formulas, despite the limited number of realizations.

In the limit $k\to 0$ (i.e., near the resonance point or if $\Delta_a,\,\Delta_{\rm pl}\ll \Delta_{a\gamma}$) the calculation of $\langle {\cal T}_{x,y}^m\rangle$ are straightforward, for example:
\begin{eqnarray}
\langle {\cal T}_x^m\rangle
&=&\left(\frac{g_{a\gamma}}{2}\right)^m\int_0^z d\zeta_1 \ldots \int_0^z d\zeta_m\, \langle\widetilde B_x^{(1)}\ldots\widetilde B_x^{(m)}\rangle\nonumber\\
&=&\left(\frac{\Delta P}{2}\right)^j\cdot\left\{\begin{array}{ll} (2j-1)!! &,\, m=2j\\
0 &,\, m=2j+1\end{array}\right.\, ,
\label{eq:phi_to_n}
\end{eqnarray}
where $(\ldots)!!$ is the double factorial and we made use of the property of the $2j$-points correlator for a gaussian field. This allows us to calculate all the moments of $p_{x,y}=|A_{x,y}|^2$. For $p_y$ we have
\begin{eqnarray}
\langle p_y^n\rangle&=&\left\langle\left[\sqrt{P_{a\gamma}^{(0)}(z)}+{\cal T}_y(z)\right]^{2n}\right\rangle\nonumber\\
&=&\sum_{j=0}^{2n}\left(\begin{array}{c}2n\\ 2j\end{array}\right)(2j-1)!!\left(\frac{\Delta P}{2}\right)^j(P_{a\gamma}^{(0)})^{n-j}\nonumber\\
&=&\sum_{j=0}^{2n}\frac{(2n)!}{j!(2n-2j)!}\left(\frac{\Delta P}{4}\right)^j (P_{a\gamma}^{(0)})^{n-j}\, .
\end{eqnarray}
We can build the moment-generating function for the distribution of $p_y$ as
\begin{eqnarray}
M_{P_y}(s)&=&{\cal L}\{f_{P_y}\}(s)=\sum_{n=0}^\infty\frac{\langle p_y^n\rangle}{n!}(-s)^n\nonumber\\
&=&\sum_{j=0}^\infty C_j\frac{(- \Delta P s/4)^j}{j!}\, .
\end{eqnarray}
where ${\cal L}$ denotes the Laplace transform, and
\begin{equation}
C_j=\sum_{n=0}^\infty\frac{(2n+2j)!}{(2n)!(n+j)!}(-P_{a\gamma}^{(0)} s)^n\, .
\end{equation}
\begin{figure}[t!]
\vspace{0.cm}
\includegraphics[width=1\columnwidth]{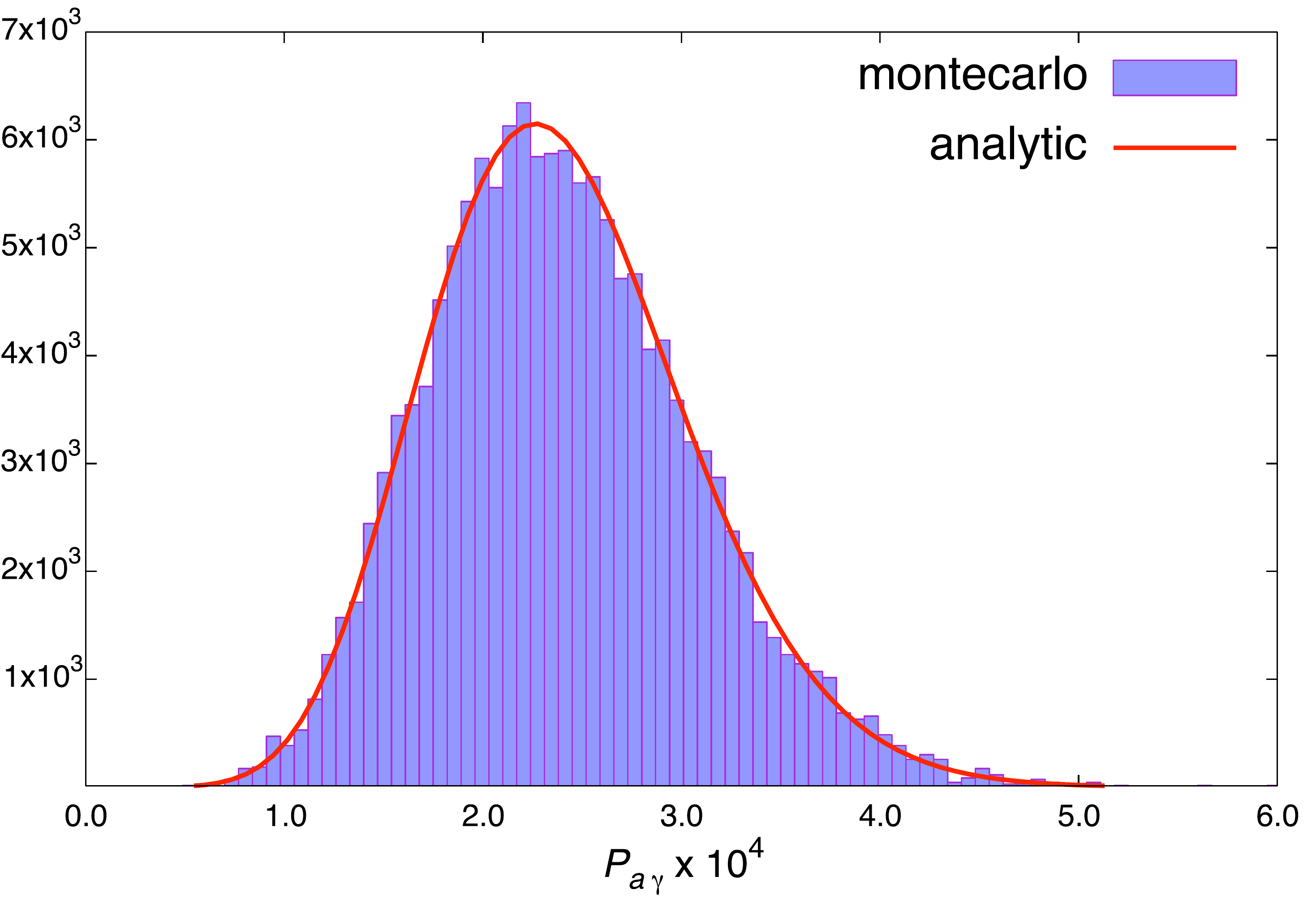}
\caption{$P_{a\gamma}$ distribution for $g_{a\gamma}=10^{-11}$~GeV$^{-1}$, $m_a=0$, $B_{\rm reg}=1$~$\mu$G, $B_{\rm rms}=1$~$\mu$G, $l_{\rm cell}=20$~pc, for a source at distance $r=1$~kpc. In blue: Monte Carlo simulation ($10^4$ realizations); In red: analytic distribution as in Eq.~(\ref{eq:PDF}).}
\label{fig:PDF1}
\end{figure}
We can prove the following identity
\begin{equation}
\sum_{n=0}^\infty(-1)^n\frac{(2n+2j)!}{(2n)!(n+j)!}t^{2n}=\frac{1}{\sqrt{\pi}}
\int_{-\infty}^{+\infty}e^{-u^2+2iut}u^{2j}\, du\, .
\label{eq:identity}\end{equation}
In fact, for $j=0$ is easily verified 
\begin{equation}
\sum_{n=0}^\infty\frac{(-1)^n}{n!}t^{2n}=e^{-t^2}=\frac{1}{\sqrt{\pi}}\int_{-\infty}^{+\infty}e^{-u^2+2iut}\, du\, .
\end{equation}
Eq.~(\ref{eq:identity}) can be proven by induction by deriving both members two times respect to the variable $t$. Using this relation we have
\begin{eqnarray}
M_{P_y}(s)&=&\frac{1}{\sqrt{\pi}}\int_{-\infty}^{+\infty}e^{-(1+\Delta Ps)u^2+2i\sqrt{P_{a\gamma}^{(0)} s}}du\nonumber\\
&=&\frac{1}{\sqrt{1+\Delta Ps}}\cdot\exp\left[\frac{-P_{a\gamma}^{(0)} s}{1+\Delta Ps}\right]\, .
\end{eqnarray}

\begin{figure}[t!]
\vspace{0.cm}
\includegraphics[width=1\columnwidth]{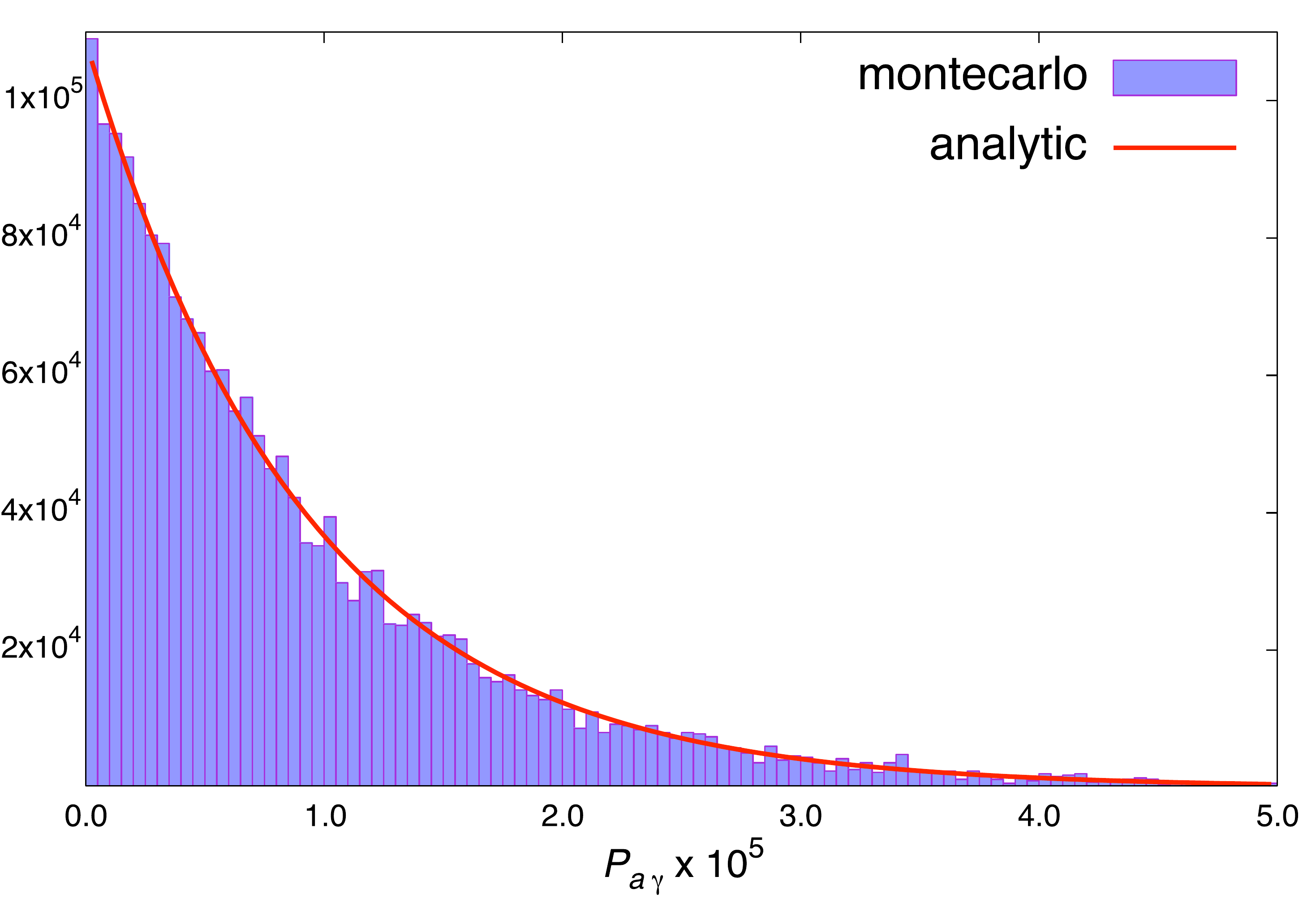}
\caption{Same as in Figure \ref{fig:PDF1} but for a pure turbulent field ($B_{\rm reg}=0$).}
\label{fig:PDF2}
\end{figure}

For $p_x$ the moment-generating function can be obtained in the same way, but with $P_{a\gamma}^{(0)}=0$. Since $p_x$ and $p_y$ are two independent variables, the probability distribution of $P_{a\gamma}=p_x+p_y$ is the convolution of $f_{P_x}$ and $f_{P_y}$. Consequently, the generating function for $P_{a\gamma}$ is just the product of the two generating functions for $p_x$ and $p_y$. Using the Bromwich integral to calculate the inverse ${\cal L}$-transform we have
\begin{eqnarray}
F(P_{a\gamma})&=&{\cal L}^{-1}\{M_{P_x}\cdot M_{P_y}\}(P_{a\gamma})\nonumber\\
&=&\frac{1}{2\pi i}\lim_{R\to\infty}\int_{\epsilon-iR}^{\epsilon+iR}
M_{P_x}(s) M_{P_y}(s)e^{sP_{a\gamma}}ds\, ,\nonumber\\
\end{eqnarray}
where $\epsilon$ is chosen in order to have the singularity $s=-1/\Delta P$ on the left of the integration path. With the change of variable $u=1+\Delta P s$ we have
\begin{equation}
F(P_{a\gamma})=\frac{e^{-\frac{P_{a\gamma}+P_{a\gamma}^{(0)}}{\Delta P}}}{2\pi i\Delta P}\lim_{R\to\infty}\int\displaylimits_{\epsilon-iR}^{\epsilon+iR}\frac{du}{u}\exp{\left(\frac{P_{a\gamma} u}{\Delta P}+\frac{P_{a\gamma}^{(0)}}{\Delta Pu}\right)}\, ,
\end{equation}
with $\epsilon>0$. The integral can be calculated closing the path on a half-circle $C_R$ in the negative real half-plane and considering that the integral on $C_R$ tends to $0$ for $R\to \infty$ due to the Jordan Lemma. The integrand has an essential singularity in $u=0$. The function
\begin{equation}
f(u)=\frac{e^{Au+B/u}}{u}\, ,
\end{equation}
can be expanded in Laurent series about $u=0$
\begin{equation}
f(u)=\sum_{n,m=0}^{\infty}\frac{A^n B^m}{n!m!}u^{n-m-1}\, ,
\end{equation}
The residue of the function is the coefficient of $u^{-1}$, that is $m=n$
\begin{equation}
\underset{u=0}{\rm Res}\, f(u)=\sum_{ n=0}^{\infty}\frac{(A B)^n}{(n!)^2}=I_0(2\sqrt{AB})\, ,
\end{equation}
where $I_0$ is the hyperbolic Bessel function of the first kind of order $0$. Making use of the residue theorem we can conclude that 
\begin{equation}
F(P_{a\gamma})=\frac{e^{-\frac{P_{a\gamma}+P_{a\gamma}^{(0)}}{\Delta P}}}{\Delta P}I_0\left(\frac{2\sqrt{P_{a\gamma}^{(0)}P_{a\gamma}}}{\Delta P}\right)\, .
\label{eq:PDF}\end{equation}
In Figs.~\ref{fig:PDF1} and \ref{fig:PDF2} we compare the $P_{a\gamma}$ distributions obtained both with a Monte Carlo simulation and with the Eq.~(\ref{eq:PDF}) for a regular field $B_{\rm reg}=1$~$\mu$G and for a pure turbulent field. The parameters used for the calculation are shown in the caption. In the last case the distribution reduces to a pure exponential, $F(P_{a\gamma})\propto e^{-P_{a\gamma}/\Delta P}$.

Finally, we notice that the previous arguments can be easily extended to the case in which ${\bf B}_{\rm reg}$ is no longer constant. In this case $P_{a\gamma}^{(0)}$ is just the oscillation probability obtained integrating Eq.~(\ref{we}) for a pure regular field (we omit the proof for simplicity).

\begin{figure}[t!]
\vspace{0.cm}
\includegraphics[width=1\columnwidth]{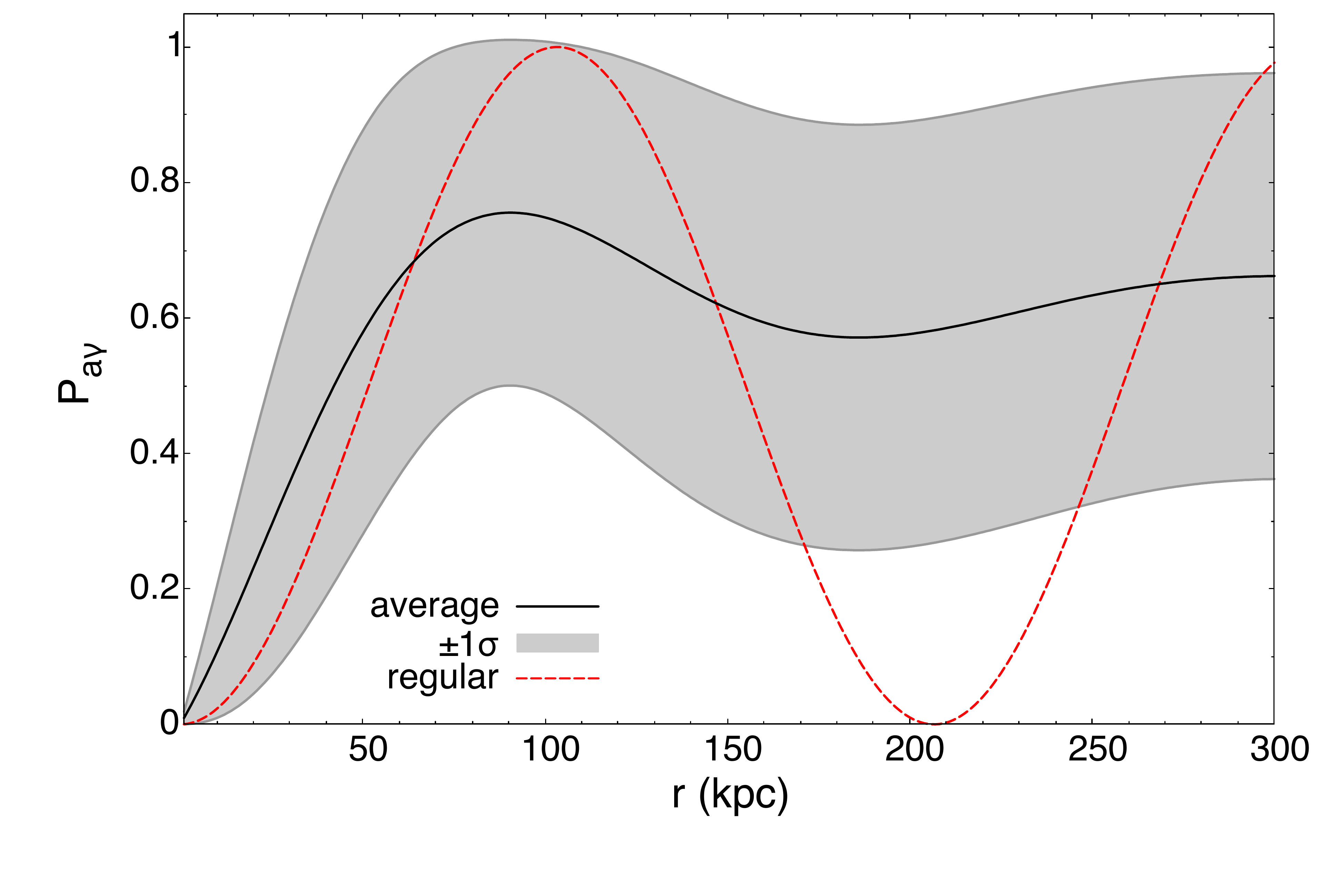}
\caption{ALP-to-photon conversion probability for $g_{a\gamma}=10^{-11}$~GeV$^{-1}$, $m_a=0$, $B_{\rm reg}=1$~$\mu$G. The red curve correspond to a regular field. The black curve and the grey band is the average of $P_{a\gamma}$ and the $\pm1\sigma$ range in presence of a turbulent field with $B_{\rm rms}=10$~$\mu$G, $l_{\rm corr}=100$~pc as calculated using Eq.~(\ref{eq:redfield2}) and its generalization.}
\label{fig:relax}
\end{figure}

\section{Gaussian turbulent component--non perturbative approach}

Previous approximations cannot be applied when $P_{a\gamma}$ is order of unity. Eq.~(\ref{we}) can be rewritten in form of Liouville equation
\begin{equation}
\frac{d\rho(z)}{dz}=-i[{\cal M}_0+\widetilde{\cal M},\rho(z)]
\label{eq:Liouville}\, ,
\end{equation}
where ${\cal M}_0$ ($\widetilde{\cal M}$) denotes the Hamiltonian containing the regular (turbulent) part of the field. However, for $\delta$--correlated gaussian perturbations is it possible to modify Eq.~(\ref{eq:Liouville}) to calculate the average of the density probability. In fact, let us write $\widetilde{\cal M}$ as
\begin{equation}
\widetilde{\cal M}=\widetilde{\Delta}_x(z){\cal Q}_x+\widetilde{\Delta}_y(z) {\cal Q}_y\, ,
\end{equation}
where $\widetilde\Delta_{x,y}=\frac{1}{2}g_{a\gamma}\widetilde B_{x,y}$ and ${\cal Q}_{a,ij}=\delta_{ia}\delta_{ja}$. With this position, it can be shown that the average matrix density $\langle\rho(z)\rangle$ satisfies the {\it Redfield} equation~\cite{Redfield,Loreti:1994ry}
\begin{equation}
\frac{d\langle\rho(z)\rangle}{dz}=-i[{\cal M}_0,\langle\rho(z)\rangle]-\beta\sum_{a=x,y}[{\cal Q}_a,[{\cal Q}_a,\langle\rho(z)\rangle]] 
\label{eq:redfield}\, ,
\end{equation}
with $\beta=\frac{1}{4}g_{a\gamma}^2B_{\rm rms}^2\l_{\rm corr}$.

To solve this equation is convenient to transform it an a ``Schr\"odinger-like'' form. 
Writing the matrix $\langle\rho_{ij}\rangle$ as a 9-component vector $R_I=\langle\rho_{ij}\rangle$, with $I=3i+j-3$ and 
$\mathbb{M}_{0,IJ}={\cal M}_{0,ik}\delta_{lj}-{\cal M}_{0,lj}\delta_{ik}$
and same for $\mathbb{Q}_{x,y}$, we can rewrite the Redfield equation as
\begin{equation}
\frac{dR(z)}{dz}=\left[-i\mathbb{M}_0-\beta(\mathbb{Q}_x^2+\mathbb{Q}_y^2)\right]R(z)\, ,
\label{eq:redfield2}\end{equation}
which has a simple formal solution (in the case of a constant ${\bf B}_{\rm reg}$ 
field)
\begin{equation}
R(z)=\exp\left[-i\mathbb{M}_0-\beta(\mathbb{Q}_x^2+\mathbb{Q}_y^2)\right]R(0)\, .
\label{eq:redfieldsol}\end{equation}
Handy subroutines for calculating exponentials of real or complex matrices can be found in the {\tt Expokit} package~\cite{expokit}. The $\langle P_{a\gamma}\rangle$ conversion probability can be calculated from the density matrix $\langle\rho\rangle$ as $\langle P_{a\gamma}\rangle=\langle\rho_{11}+\rho_{22}\rangle$ with the initial condition $\langle\rho_{ij}(0)\rangle=\delta_{i3}\delta_{j3}$.

Higher order moments of $P_{a\gamma}$ can be calculated by generalizing Eq.~(\ref{eq:redfield2}) to tensorial products of the matrix $\rho$. For example, for the second momentum we consider $\rho^{(2)}=\rho\otimes\rho$. Using the 81-component vector $R^{(2)}=R\otimes R$, that is $R^{(2)}_{\cal I}=R_IR_J$, 
${\cal I}=9I+J-9$ and 
$\mathbb{M}^{(2)}_{0,{\cal IJ}}=\mathbb{M}_{0,IJ}\delta_{lKL}+\mathbb{M}_{0,KL}\delta_{IJ}$
and so on, repeating the same argument used to obtain Eq.~(\ref{eq:redfield2}), it can be shown that the Redfield equation for the tensor product have the same form of Eq.~(\ref{eq:redfield2}) and thus the same formal solution of Eq.~(\ref{eq:redfieldsol}) (we omit here the proof). The variance of $P_{a\gamma}$ can be calculated as $\langle P_{a\gamma}^2\rangle=\langle(\rho_{11}+\rho_{22})^2\rangle=\langle\rho^{(2)}_{11,11}+\rho^{(2)}_{22,22}+2\rho^{(2)}_{11,22}\rangle$. Although the same argument can be used to calculate any moment of $P_{a\gamma}$ the complexity of Eq.~(\ref{eq:redfield2}) grows exponentially. 

In Figure \ref{fig:relax} we show the ALP-to-photon conversion probability for a regular field $B_{\rm reg}=1$~$\mu$G (red curve) and with a turbulent field $B_{\rm rms}=10$~$\mu$G and a correlation length $l_{\rm cell}=100$~pc (black curve) together with the $\pm 1\sigma$ band calculated with the help of Eq.~(\ref{eq:redfield2}). The value of the turbulent field and the correlation length as well as the $z$ range chosen for this plot are unrealistic for the Milky-Way and are intended only for illustrative purposes. Notice that the distribution of $P_{a\gamma}$ is neither gaussian nor symmetric in general, so that the gray band should to be intended just as a qualitative range without a defined confidence level (and in fact sometimes the band comes out of the range $[0,1]$). 
We notice the typical effect induced by stochastic term in the Hamiltonian, i.e., the flavor composition tends to be equally distributed among all the degree of freedom, and thus $\langle P_{a\gamma}\rangle\to 2/3$ for $z\to\infty$. This phenomenon is well known for example in neutrino oscillations in presence of a dissipative term (see, e.g., \cite{Barenboim:2004wu}).

\end{document}